\documentclass[acmsmall, nonacm]{acmart}
\setcopyright{none}
\makeatletter
\renewcommand\footnotetextcopyrightpermission[1]{}
\renewcommand\footnotetextauthorsaddresses[1]{}
\makeatother

\usepackage{booktabs} 
\usepackage[ruled]{algorithm2e} 
\usepackage{natbib} 

\SetAlFnt{\small}
\SetAlCapFnt{\small}
\SetAlCapNameFnt{\small}
\SetAlCapHSkip{0pt}
\IncMargin{-\parindent}

\usepackage{amsmath}

\usepackage{amssymb}
\usepackage{amsthm}
\usepackage{graphicx}
\usepackage{subcaption}
\usepackage{caption}
\usepackage{multirow}
\usepackage{xcolor}
\usepackage{longtable}
\usepackage{mathtools}
\usepackage{float}

\setcitestyle{authoryear}

\newtheorem{assumption}{Assumption}

\newtheorem{theorem}{Theorem}
\newtheorem{lemma}{Lemma}
\newtheorem{corollary}{Corollary}
\newtheorem{definition}{Definition}
\newtheorem{remark}{Remark}

\newcommand{\sC}{s_C^*}
\newcommand{\sT}{s_T^*}
\newcommand{\hatsC}{\hat{s}_C}
\newcommand{\hatsT}{\hat{s}_T}
\newcommand{\aL}{\alpha_L}
\newcommand{\aC}{\alpha_C}
\newcommand{\eps}{\varepsilon}
\newcommand{\TTE}{\mathrm{GTE}}
\newcommand{\GTE}{\widehat{\mathrm{GTE}}}
\newcommand{\BiasDD}{\mathrm{Bias}_{\mathrm{DD}}}
\newcommand{\BiasTSRN}{\mathrm{Bias}_{\mathrm{TSRN}}}

\newcommand{\dd}{\mathrm{d}}

\title[The Double Diagonal Estimator for Two-Sided Marketplace Experiments]{The Double Diagonal Estimator: Reducing Bias in Two-Sided Marketplace Experiments}

\author{Saanvi Chawla}
\email{saanvic@stanford.edu}
\orcid{0009-0004-7184-6908}
\affiliation{%
  \institution{Stanford University}
  \city{Stanford}
  \state{CA}
  \country{USA}
}

\author{Vasilis Syrgkanis}
\email{vsyrgk@stanford.edu}
\orcid{0000-0001-5047-8772}
\affiliation{%
  \institution{Stanford University}
  \city{Stanford}
  \state{CA}
  \country{USA}
}

\begin{abstract}
Interference between participants in two-sided marketplace experiments can severely bias conventional A/B tests. While two-sided randomization (TSR), which randomizes both supply and demand, enables estimators that mitigate this bias, existing analyses offer limited guidance on which estimator to prefer for the balanced market conditions common in practice, as they largely focus on extreme supply- or demand-constrained regimes. We introduce the Double Diagonal (DD) estimator, a novel method for interference-robust treatment effect estimation. The DD estimator contrasts the outcomes of treated-treated and control-control pairs and uses the off-diagonal interactions (treated-control and control-treated) to correct for interference-induced bias. We provide the first theoretical comparison between the DD and the standard naive TSR (TSRN) estimator under a simplifying mean-field market model, proving two key results: (i) DD's bias is strictly smaller than the standard estimator's below a critical threshold of the outside-option parameter; in particular, DD's bias is smaller whenever the outside option is at most twice as attractive as platform options; and (ii) the two estimators exhibit biases in opposite directions, with DD underestimating and the standard estimator overestimating the true effect, respectively. Extensive numerical experiments in realistic market settings—spanning heterogeneous treatment effects, supply-demand imbalances, and varied market structures—corroborate our theoretical findings. Across a wide range of configurations, the DD estimator demonstrates the lowest bias, establishing it as a robust and often superior choice for reliable experimentation in two-sided marketplaces.
\end{abstract}

\begin{document}

\makeatletter
\fancypagestyle{standardpagestyle}{%
  \fancyhf{}%
  \fancyfoot[C]{\thepage}%
  \renewcommand{\headrulewidth}{0pt}%
  \renewcommand{\footrulewidth}{0pt}%
}
\fancypagestyle{firstpagestyle}{%
  \fancyhf{}%
  \fancyfoot[C]{\thepage}%
  \renewcommand{\headrulewidth}{0pt}%
  \renewcommand{\footrulewidth}{0pt}%
}
\pagestyle{standardpagestyle}
\makeatother

\maketitle

\section{Introduction}

Two-sided platforms, including lodging marketplaces like Airbnb, freelancing platforms like Upwork, and ride-sharing services like Uber, routinely employ A/B testing to evaluate interventions that affect interactions between buyers and sellers. The fundamental challenge in these experiments is \emph{interference}: the treatment assigned to one market participant influences outcomes for other participants through competition for scarce resources. When a seller's listing becomes more attractive due to treatment, it may ``steal'' bookings from control listings, inflating the apparent treatment effect. This violation of the Stable Unit Treatment Value Assumption (SUTVA) \citep{imbens2015causal} renders naive estimators biased, sometimes by factors of two or more \citep{blake2014marketplace}.

Prior work has established that the direction and magnitude of bias depend critically on market balance. \citet{johari2021experimentaldesigntwosidedplatforms} developed a mean field framework showing that customer-side randomization (CR) is unbiased in highly demand-constrained markets while listing-side randomization (LR) is unbiased in highly supply-constrained markets. They introduced Two-Sided Randomization (TSR), which randomizes both customers and listings simultaneously, along with a naive TSR estimator (TSRN). While TSR represents an important advance, the question of optimal estimator design remains open: Can we do better than existing approaches?

\paragraph{Our Contributions.} We introduce the \emph{Double Diagonal} (DD) estimator and provide rigorous theoretical and empirical evidence that it achieves lower bias than TSRN across a wide range of market conditions:

\begin{enumerate}
    \item \textbf{A novel estimator.} We propose the Double Diagonal estimator, which contrasts the outcomes of treated-treated and control-control pairs and uses off-diagonal interactions to correct for interference-induced bias.
    
    \item \textbf{Theoretical bias characterization.} We prove that for small treatment effects: (i) DD systematically \emph{underestimates} the true treatment effect ($\BiasDD < 0$), while (ii) TSRN systematically \emph{overestimates} ($\BiasTSRN > 0$).
    
    \item \textbf{Explicit optimality threshold.} We prove the existence of a unique critical threshold $a^*(\aL)$ such that $|\BiasDD| < |\BiasTSRN|$ for all $a < a^*(\aL)$. At $\aL = \frac{1}{2}$, we derive $a^* \approx 2.34$, meaning DD's bias is smaller whenever the outside option is at most about twice as attractive as platform options.
    
    \item \textbf{Experimental validation.} Extensive numerical experiments across diverse market configurations corroborate our theoretical findings, with DD achieving the lowest bias in most scenarios.
\end{enumerate}

\paragraph{Organization.} Section~\ref{sec:model} presents the mean field market model. Section~\ref{sec:experiment} describes two-sided experimental design. Section~\ref{sec:estimators} defines the estimators. Section~\ref{sec:theory} contains our main theoretical results. Section~\ref{sec:experiments} presents experimental evaluation. Section~\ref{sec:conclusion} concludes.

\section{Related Work}\label{sec:rel_work}

\citet{johari2021experimentaldesigntwosidedplatforms} develop an analytical framework for experimental design in two-sided platforms using a mean-field market model. They show that the bias of standard one-sided estimators depends critically on market balance: customer-side randomization (CR) is unbiased in highly demand-constrained markets, while listing-side randomization (LR) is unbiased in highly supply-constrained markets. They introduce two-sided randomization (TSR), where both customers and listings are independently assigned to treatment and control, and propose a naive TSR estimator (TSRN) along with improved interpolating estimators that heuristically correct for interference. Our work builds directly on this framework: we adopt the same mean-field market model and TSR experimental design, but ask a different question—given that two-sided randomization is employed, which estimator should the platform use? We introduce the Double Diagonal (DD) estimator, which aggregates bookings by listing cohort rather than by customer-listing combination, and provide the first theoretical comparison between DD and TSRN. While Johari et al. characterize bias in the limiting regimes of extreme market imbalance, our results address the practically important intermediate regime, proving that DD has strictly lower bias than TSRN below a critical threshold of the outside-option parameter and that the two estimators exhibit biases in opposite directions.

\citet{masoero2026multiple} provide the first formal finite-sample analysis of Multiple Randomization Designs (MRDs), a general class of experimental designs for marketplaces with multiple interacting populations. Under a local interference assumption, where the outcome of a buyer-seller pair depends only on its own treatment assignment and the fractions of treated buyers and sellers, they derive exact finite-sample variances for plug-in estimators of causal effects in Simple MRDs (SMRDs), propose conservative variance estimators, and prove design-based central limit theorems. Their analysis is design-based and nonparametric, placing no distributional assumptions on potential outcomes, and they show that SMRDs can detect and measure spillover effects that single-sided designs cannot identify. Our work is complementary: while Masoero et al. focus on variance characterization and inference under local interference, we focus on bias characterization under a parametric mean-field model and analyze bias with respect to Global Treatment and Control. The DD estimator we propose can be expressed within the SMRD framework of Masoero et al., where it corresponds to a specific linear combination of the four cohort-level averages, and the finite-sample inference tools they develop could in principle be applied to conduct valid inference for DD. The MRD framework is also discussed in the review by \citet{bajari2023experimental}, who introduce MRDs and discuss their application to marketplace experimentation more broadly.

Several recent papers study interference and bias correction specifically in two-sided marketplace experiments. \citet{li2022interference} develop a tractable market model to study how the choice of experiment side (supply vs. demand) and treatment proportion affect both bias and variance, providing practical guidance for platforms. \citet{bright2025reducing} propose using shadow prices from the platform's matching linear program to reduce interference bias, proving that their estimator corresponds to the correct first-order Taylor approximation of the value function and is less biased than the standard RCT estimator. \citet{farias2023correcting} introduces a Differences-in-Qs Monte-Carlo estimator that formalizes interference correction as a policy evaluation problem and achieves bias that is second-order in the treatment effect. \citet{nandy2021b} propose the UniCoRn approach, a producer-side experiment design for two-sided marketplaces that is agnostic to network density and does not require prior knowledge of the underlying network. \citet{goli2024bias} propose a bias-correction approach that leverages data across multiple experiments to identify observations near partial equilibrium. Our work differs from these approaches in that we do not change experimental design or introduce a post-hoc bias correction method; instead, we introduce a new estimator for the existing TSR design and provide theoretical guarantees on when it dominates the standard alternative.

 Foundational work on interference in randomized experiments includes \cite{rosenbaum2007interference, hudgens2008toward, manski2013identification, ugander2013graph, eckles2017design}, which developed frameworks for understanding how treatment spillovers violate the stable unit treatment value assumption (SUTVA). More recently, researchers have focused specifically on two-sided marketplace settings, where interference arises from equilibrium effects between supply and demand \citep{johari2024does, wager2021experimenting, bajari2023experimental, munro2025treatment, shirani2024causal}. Several studies have characterized the bias in estimating global treatment effects (GTE) from platform experiments when treatments are sign-consistent \citep{hu2022average, aronow2017estimating, han2023detecting, johari2017peeking}. We contribute to this literature by providing a theoretical comparison between competing TSR estimators under balanced market conditions and deriving conditions under which different estimators are preferred based on the market's outside-option parameter. 

Finally, a related line of work studies interference through the lens of bipartite experimental designs. \citet{doudchenko2020causal} and \citet{zigler2020bipartite} develop causal inference methods for bipartite experiments, where treatment is applied to one set of units (e.g., sellers) and outcomes are measured on a different set (e.g., buyers). Doudchenko et al. leverage the generalized propensity score literature to obtain unbiased estimates of causal effects for bipartite experiments and construct confidence sets with proper coverage probabilities. These bipartite designs are particularly natural in marketplace settings where assigning treatment on one side and measuring outcomes on the other can better account for cross-side interference. While this line of work models interference through a fixed bipartite graph structure connecting two populations, our setting captures interference that arises endogenously through dynamic competition for limited inventory, where the interference pattern itself depends on the experimental design and market conditions.

\section{The Mean Field Market Model}\label{sec:model}

We adopt the mean field market model introduced by \citet{johari2021experimentaldesigntwosidedplatforms}. This section provides a complete description of the model primitives, dynamics, and steady-state characterization.

\subsection{Model Primitives}

Consider a marketplace with $n$ customer types and $m$ listing types operating in continuous time.

\paragraph{Listings.} There is a unit mass of listings distributed across types according to $\rho = (\rho_1, \ldots, \rho_m)$ where $\sum_j \rho_j = 1$. The \emph{state} of the market is the vector $s = (s_1, \ldots, s_m)$ representing the mass of available (unoccupied) listings of each type. From the mass $\rho_j - s_j$ of occupied listings of type $j$, a fraction $\tau_j$ returns to the market per unit time.

\paragraph{Customers.} A mass $\mu$ of customers arrives per unit time, with types distributed according to $\phi = (\phi_1, \ldots, \phi_n)$ where $\sum_i \phi_i = 1$. Each arriving customer of type $i$ chooses among available listings according to the choice probability:
\begin{equation}\label{eq:choice_prob}
\pi_{ij}(s) = \frac{P_{ij} \cdot s_j}{\epsilon_i + \sum_{k} P_{ik} \cdot s_k},
\end{equation}
where $P_{ij} \geq 0$ is the \emph{matching propensity} between customer type $i$ and listing type $j$, and $\epsilon_i > 0$ is the \emph{outside option} value for customer type $i$ (representing the utility of not booking).

\begin{remark}[Microfoundation]
The choice probability \eqref{eq:choice_prob} arises from an attention-logit model: each listing of type $j$ enters customer $i$'s consideration set with probability $\alpha_{ij}$, and the customer chooses proportionally to values $v_{ij}$ among considered listings. With $s_j$ available listings, the expected consideration set contains $\alpha_{ij} s_j$ listings of type $j$, yielding \eqref{eq:choice_prob} with $P_{ij} = v_{ij} \alpha_{ij}$.
\end{remark}

\subsection{Platform Dynamics}

The market state evolves according to the balance between listings returning to availability and listings being booked. In the infinitesimal time interval $[t, t + \eta]$:
\begin{equation}
s_{t+\eta} = s_t + \eta \cdot \frac{ds_t}{dt},
\end{equation}
where the rate of change comprises two components:
\begin{enumerate}
    \item \textbf{Inflow:} Occupied listings of type $j$ return at rate $(\rho_j - s_j) \tau_j$.
    \item \textbf{Outflow:} Available listings of type $j$ are booked at rate $\mu \sum_i \phi_i \pi_{ij}(s)$.
\end{enumerate}
The dynamics are therefore:
\begin{equation}\label{eq:dynamics}
\frac{ds}{dt} = (\rho - s) \cdot \tau - \mu \sum_i \phi_i \cdot \pi_i(s),
\end{equation}
where $\pi_i(s) = (\pi_{i1}(s), \ldots, \pi_{im}(s))$ is the vector of choice probabilities for customer type $i$.

\subsection{Steady-State Equilibrium}

The dynamical system \eqref{eq:dynamics} admits a unique steady state $s^*$ characterized by zero net flow:
\begin{equation}\label{eq:steady_state}
(\rho - s^*) \cdot \tau = \mu \sum_i \phi_i \cdot \pi_i(s^*).
\end{equation}
The left side represents the rate at which listings become available; the right side represents the rate at which they are booked. Note that for a homogeneous turnover rate $\tau$, the steady state $s^*$ only depends on the ratio $\mu / \tau$. Hence, without loss of generality, we can normalize the model such that $\tau=1$. For any general $(\tau, \mu)$ model, we can always analyze the rescaled model $(1, \mu/\tau)$. Such a model will have the same steady state equilibrium $s^*$ and the booking rate of every listing type will simply be a scaled version of the booking rate under the original model. Such uniform re-scalings of the booking rates across listing types will not affect any of our downstream analyses. Hence, in the remainder of the paper we make the following simplifying convention:
\begin{assumption}[Normalization, without loss of generality] We assume that $\tau=1$.
\end{assumption}

\subsection{Platform Objective and Global Treatment Effect}

The platform's objective is to maximize the steady-state booking rate:
\begin{align}\label{eq:booking_rate}
Q =~& \sum_{j=1}^m Q_j, & Q_j \coloneqq~& \mu \sum_i \phi_i \cdot \pi_{ij}(s^*).
\end{align}
Any intervention, such as displaying a superhost badge or modifying search rankings, affects the matching propensity matrix $P$, which in turn affects the steady state $s^*$ and hence the booking rate $Q$. Consider an intervention that changes the matching matrix from $P$ (control) to $\tilde{P}$ (treatment).

\begin{definition}[Global Treatment Effect]
Let $s^*_{\text{control}}$ denote the steady state under $P$ and $s^*_{\text{treatment}}$ the steady state under $\tilde{P}$. The \emph{Global Treatment Effect} (GTE) is:
\begin{equation}\label{eqn:gte}
\text{GTE} = Q_{\text{treatment}} - Q_{\text{control}},
\end{equation}
where $Q_{\text{treatment}}$ and $Q_{\text{control}}$ are the booking rates at the respective steady states.
\end{definition}

\noindent The GTE represents the causal effect of deploying the treatment globally to all customers and listings on the platform. It will also be convenient for our analysis to relate it to the steady state availability of listings.

\begin{lemma}
    Under the mean field market model the GTE is equal to the difference between steady state available inventory under control, vs. steady state available inventory under treatment:
    \begin{align}\label{eqn:gte-alt}
        GTE = \sum_{j=1}^m s_{\text{control}, j}^{*} - \sum_{j=1}^m s_{\text{treatment},j}^{*}
    \end{align}
    i.e., it measures how many less total listings remain available on the platform in steady state.
\end{lemma}
\begin{proof}
    Note that by the Steady-State equilibrium condition (with the normalization $\tau=1$), we have that under global treatment and global control correspondingly, for any listing type $j\in [m]$:
    \begin{align*}
        \rho_j - s_{\text{treatment}, j}^{*} =~& Q_{\text{treatment},j} &
        \rho_j - s_{\text{control}, j}^{*} =~& Q_{\text{control}, j}
    \end{align*}
    Subtracting the two equations and summing across $j\in [m]$ yields the result.
\end{proof}

\section{Two-Sided Randomized Experiments}\label{sec:experiment}

Rather than deploying treatment globally, platforms conduct experiments to estimate the GTE before committing to a policy change. A \emph{Two-Sided Randomized Experiment} (TSRE) randomizes both customers and listings:
\begin{itemize}
    \item Each customer is assigned to treatment with probability $\aC$ and control with probability $1 - \aC$.
    \item Each listing is assigned to treatment with probability $\aL$ and control with probability $1 - \aL$.
    \item The new technology $\tilde{P}$ is applied only when a treatment customer considers a treatment listing; all other interactions use the control technology $P$.
\end{itemize}

\subsection{Experimental Dynamics}

Under the TSRE, we can model the system by simply introducing ``doubled'' type spaces: customer types $(i, d)$ for $i \in [n]$, $d \in \{0, 1\}$ and listing types $(j, z)$ for $j \in [m]$, $z \in \{0, 1\}$, where $d, z = 1$ indicates treatment assignment. In this doubled type space, the experimental matching matrix is:
\begin{equation}
P_e = \begin{pmatrix} P & P \\ P & \tilde{P} \end{pmatrix},
\end{equation}
where rows correspond to customer cohorts (control, treatment) and columns to listing cohorts. The arrival rates and masses of these doubled types become:
\begin{align}
\phi_{i,0} &= \phi_i (1 - \aC), & \phi_{i,1} &= \phi_i \aC, \\
\rho_{j,0} &= \rho_j (1 - \aL), & \rho_{j,1} &= \rho_j \aL.
\end{align}

\subsection{Booking Rate Decomposition}

Under the experiment, let $s^*_e$ denote the steady state. The total booking rate decomposes across cohort pairs:
\begin{equation}
Q = \begin{pmatrix} Q_{00} & Q_{01} \\ Q_{10} & Q_{11} \end{pmatrix},
\end{equation}
where $Q_{dz}$ is the booking rate from customer cohort $d$ to listing cohort $z$:
\begin{equation}
Q_{dz} = \mu \sum_i \phi_{(i,d)} \sum_j \pi_{(i,d),(j,z)}(s^*_e).
\end{equation}

\section{Treatment Effect Estimators}\label{sec:estimators}

An estimator of the GTE from a two-sided randomized experiment is any function of the observed steady state booking rates between each listing type $(j, z)$ and customer type pair $(i,d)$. A natural class of estimators also considered in the seminal works of \citet{johari2021experimentaldesigntwosidedplatforms,masoero2026multiple} on two-sided randomization are linear combinations of the four observable cohort-level booking rates $Q_{00}$, $Q_{01}$, $Q_{10}$, and $Q_{11}$. We will restrict our investigation to such estimators. The central question of our work is: \emph{which linear combination has the smallest bias relative to the true GTE under the mean field structural model?} 

In this section, we review several estimators proposed in prior work and which have only been analyzed theoretically, in terms of bias, in extreme market regimes. In Section~\ref{sec:dd}, we introduce the Double Diagonal (DD) estimator, our proposed answer to this question. 

\subsection{The Naive TSR (TSRN) Estimator}

The Naive Two-Sided Randomization estimator \citep{johari2021experimentaldesigntwosidedplatforms} treats the $(1,1)$ cell as representative of global treatment:
\begin{equation}\label{eq:TSRN}
\GTE_{\text{TSRN}} = \frac{Q_{11}}{\aC \cdot \aL} - \frac{Q_{01} + Q_{10} + Q_{00}}{1 - \aC \cdot \aL}.
\end{equation}
The intuition is that only treated customer--treated listing pairs experience the treatment technology, while all other pairs experience control.

\subsection{One-Sided Estimators}

One-sided estimators condition on the subset of treated units on one side of the market and look at the difference in booking rates on the other side:
\begin{align}
\GTE_{\text{CR}} &= \frac{Q_{11}}{\aC \cdot \aL} - \frac{Q_{01}}{(1-\aC) \cdot \aL} \quad \text{(Customer Randomization)}, \\
\GTE_{\text{LR}} &= \frac{Q_{11}}{\aC \cdot \aL} - \frac{Q_{10}}{\aC \cdot (1-\aL)} \quad \text{(Listing Randomization)}.
\end{align}

\subsection{The TSRI-k Estimators}

\cite{johari2021experimentaldesigntwosidedplatforms} introduce the TSRI-k (Two-Sided Randomization Improved) family of estimators, which incorporate corrections for interference effects by interpolating between customer-side and listing-side perspectives based on market balance. Define the normalized cell booking rates:
\begin{equation}
\bar{Q}_{ij} = \frac{Q_{ij}}{a_C^{[i=1]} (1-a_C)^{[i=0]} \cdot a_L^{[j=1]} (1-a_L)^{[j=0]}},
\end{equation}
where $[i=1]$ denotes the indicator function.  The TSRI-k estimator is:
\begin{align}
\widehat{\text{GTE}}_{\text{TSRI-}k} &= \beta \left[ \bar{Q}_{11} - \bar{Q}_{01} - k(1-\beta)(\bar{Q}_{00} - \bar{Q}_{01}) \right] \nonumber \\
&\quad + (1-\beta) \left[ \bar{Q}_{11} - \bar{Q}_{10} - k\beta(\bar{Q}_{00} - \bar{Q}_{10}) \right].
\end{align}
The correction terms $(\bar{Q}_{00} - \bar{Q}_{01})$ and $(\bar{Q}_{00} - \bar{Q}_{10})$ capture interference effects: the former measures how control customers' booking rates differ between control and treatment listings (listing-side cannibalization), while the latter measures how booking rates on control listings differ between control and treatment customers (customer-side competition). The parameter $k$ controls the strength of these corrections, with $k=1$ (TSRI-1) and $k=2$ (TSRI-2) being the primary variants studied. TSRI-2 achieves lower bias but higher variance than TSRI-1. The interpolation weight $\beta = e^{-\mu/\tau}$ depends on the ratio of customer arrival rate $\mu$ to listing turnover rate $\tau$: when $\mu/\tau \to 0$ (demand-constrained), $\beta \to 1$ and the estimator emphasizes the CR component; when $\mu/\tau \to \infty$ (supply-constrained), $\beta \to 0$ and the estimator emphasizes the LR component.

\section{The Double Diagonal Estimator} \label{sec:dd}

Our quest for a good candidate estimator begins in realizing the asymmetry of interference in this mean field market model: listings are depletable inventory but customers are a flow. When a treatment listing gets booked, it becomes unavailable to all future customers - both treated and control - until it replenishes.  This availability reduction affects all subsequent customers. Customer treatment status, by contrast, affects only that individual's booking decision without depleting any shared resource. Treatment listings with enhanced attractiveness get booked faster, depleting their availability; this availability reduction is the treatment effect operating through the supply side. 

Even though treated listings are sometimes showcased without the treatment technology, in balanced market conditions with substantial inter-temporal interference, we expect that the steady state availability of treated listings will resemble their steady state availability under global treatment in a market with a mass $\aL$ of listings (especially in settings where the effect of the technology is small). Similarly, given that control listings are never displayed with the treatment technology, we expect the steady state availability of control listings to resemble their steady state availability under global control in a market with a mass $1-\aL$ of listings.

Given this asymmetry, instead of trying to approximate the primary definition of the GTE given in Equation~\eqref{eqn:gte}, we will instead attempt to approximate the alternative equivalent definition given in Equation~\eqref{eqn:gte-alt}. The GTE is the difference in available (global) control vs (global) treated listings at steady state. We will consider an estimator which matches this definition and simply drops the \emph{global}: we will look for an estimator that, even under the experimental dynamics, measures the difference between the available control listings vs. the available treated listings at steady state (each rescaled by their corresponding population size to simulate the global market): 
\begin{align}
        \frac{1}{1-\aL} \sum_{j=1}^m s_{(j,0)}^* - \frac{1}{\aL} \sum_{j=1}^m s_{(j,1)}^* 
\end{align}
These steady state occupancies implicitly incorporate interference patterns as discussed above. The next Lemma shows that this quantity can be written as an estimator that corresponds to a particular linear combination of the four observable cohort-level booking rates $Q_{00}$, $Q_{01}$, $Q_{10}$, and $Q_{11}$ that we will refer to as the \emph{Double Diagonal} estimator.

\begin{lemma}\label{lem:equiv-form}
    Under the mean field market model the normalized difference in experimental steady state availabilities between control and treated listings can be written as the following estimator:
    \begin{align}
        \frac{1}{1-\aL} \sum_{j=1}^m s_{(j,0)}^* - \frac{1}{\aL} \sum_{j=1}^m s_{(j,1)}^* = \frac{Q_{11} + Q_{01}}{\aL} - \frac{Q_{10} + Q_{00}}{1-\aL}
    \end{align}
    We will refer to the right hand side as the \emph{Double Diagonal} (DD) estimator:
\begin{equation}\label{eq:DD}
\GTE_{\text{DD}} = \frac{Q_{11} + Q_{01}}{\aL} - \frac{Q_{10} + Q_{00}}{1 - \aL}.
\end{equation}
\end{lemma}
\begin{proof}
    The steady state equilibrium constraints dictate that:
    \begin{align*}
    (1 - \aL) \rho_{j} - s_{(j,0)}^* =~& Q_{0, (j,0)} + Q_{1, (j,0)} &
    \aL \rho_{j} - s_{(j,1)}^* =~& Q_{0, (j,1)} + Q_{1, (j, 1)}
    \end{align*}
    where:
    \begin{align*}
        Q_{0, (j, z)} \coloneqq~& (1 - \aC) \mu\sum_{i=1}^n \phi_i \pi_{(i,0),(j,z)} &
        Q_{1, (j, z)} \coloneqq~& \aC \mu\sum_{i=1}^n \phi_i \pi_{(i,1),(j,z)} 
    \end{align*}
    Adding the first two equations across $j$, we have:
    \begin{align*}
    (1 - \aL) - \sum_{j=1}^m s_{(j,0)}^* =~& Q_{00} + Q_{10} &
    \aL - \sum_{j=1}^m s_{(j,1)}^* =~& Q_{01} + Q_{11}
    \end{align*}
    Dividing the first equation by $1-\aL$ and the second equation by $\aL$:
    \begin{align*}
    1 - \frac{1}{1 - \aL}\sum_{j=1}^m s_{(j,0)}^* =~& \frac{Q_{00} + Q_{10}}{1 - \aL} &
    1 - \frac{1}{\aL}\sum_{j=1}^m s_{(j,1)}^* =~& \frac{Q_{01} + Q_{11}}{\aL}
    \end{align*}
    Subtracting the two equations, we derive:
    \begin{align*}
        \frac{1}{1-\aL} \sum_{j=1}^m s_{(j,0)}^* - \frac{1}{\aL} \sum_{j=1}^m s_{(j,1)}^* = \frac{Q_{11} + Q_{01}}{\aL} - \frac{Q_{10} + Q_{00}}{1-\aL}
    \end{align*}
\end{proof}

\noindent The DD estimator compares total bookings of treatment listings against total bookings of control listings, aggregating by listing cohort rather than by customer-listing combination. From another viewpoint, the DD estimator contrasts the outcomes of treated-treated and control-control pairs (the "diagonal" terms $Q_{11}$ and $Q_{00}$) while using the off-diagonal interactions ($Q_{01}$ and $Q_{10}$) to correct for interference-induced bias. From this perspective, the estimator also performs some interference bias correction. This interference bias correction can be made more explicit if we express the DD estimator in terms for the fully normalized booking rates $\bar{Q}_{ij}$ that we used to define the TSRI-$k$ estimators considered in the prior work of \citet{johari2021experimentaldesigntwosidedplatforms}:
\begin{align}
     \frac{Q_{11} + Q_{01}}{\aL} - \frac{Q_{10} + Q_{00}}{1-\aL} =~& \aC \bar{Q}_{11} + (1 - \aC) \bar{Q}_{01} - \aC \bar{Q}_{10} - (1-\aC)\bar{Q}_{00} \nonumber\\
     =~& \bar{Q}_{11} - (1 - \aC) \bar{Q}_{11} + (1 - \aC) \bar{Q}_{01} - \aC (\bar{Q}_{10} - \bar{Q}_{00}) - \bar{Q}_{00}\nonumber\\
     =~& \underbrace{\bar{Q}_{11} - \bar{Q}_{00}}_{\text{Diagonal Effect}} - (1 - \aC) \underbrace{(\bar{Q}_{11} - \bar{Q}_{01})}_{\text{CR effect}} - \aC \underbrace{(\bar{Q}_{10} - \bar{Q}_{00})}_{\text{Spillover effect}} \label{eqn:interp-dd}
\end{align}
The diagonal estimator calculates the diagonal effect between (treated, treated) and (control, control) pairs, then removes part of that effect that can be attributed to treated listings being exposed without the innovation to control customers. It also removes another part that can be attributed to spillover effects due to control listings competing with treated listings when exposed to treated customers.

\begin{figure}[h!]
\centering
\includegraphics[width=.7\textwidth]{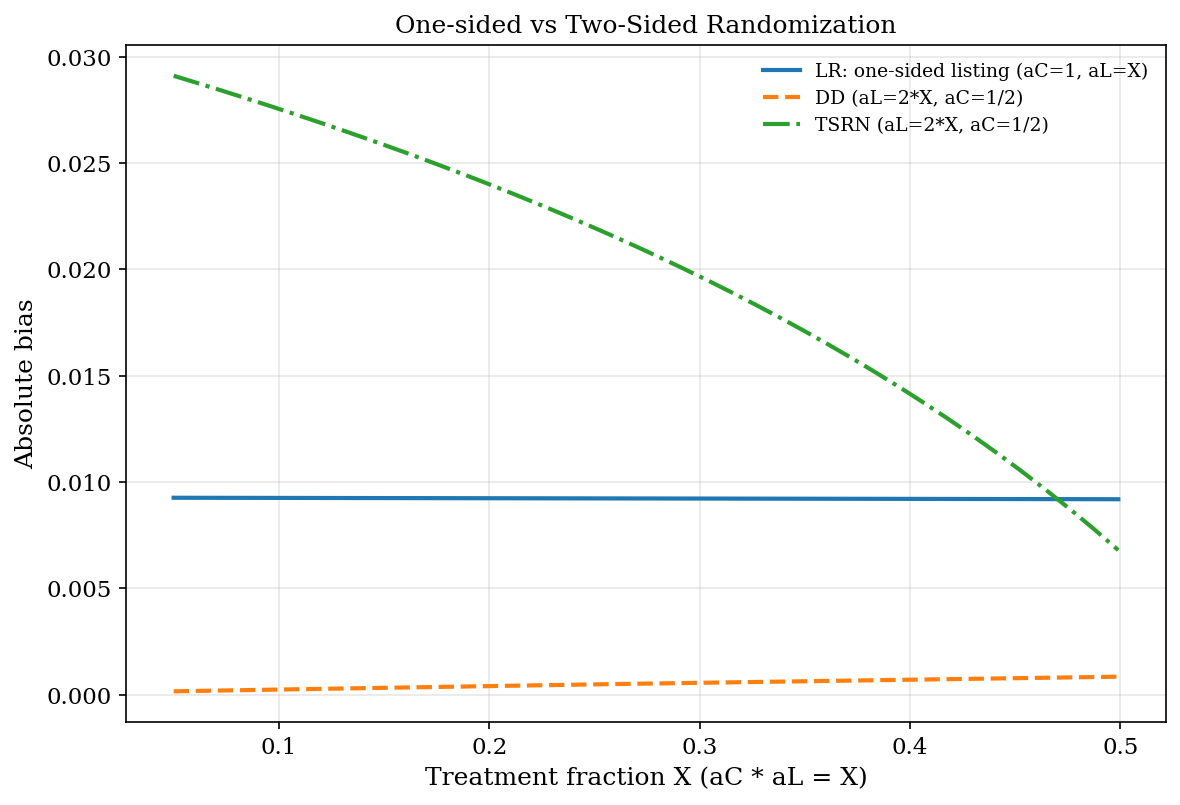}
\caption{Bias across one sided randomization for LR and two sided for DD and TSR. DD achieves minimum bias consistently.}
\label{fig:onetwo}
\end{figure}

These correction terms carry important interference bias information. Given the description of the DD estimator as difference in booking rates between treated listings vs. control listings in a two-sided randomized experiment, one might wonder: would it not then make sense to simply run one-sided listing randomization (i.e. set $\aC=1$) and measure the effect between treated and control listings? Wouldn't that also have a similarly small bias as the DD estimator? As we depict in Figure~\ref{fig:onetwo}, such an approach would have a larger bias in measuring the treatment effect. Intuitively, such an approach does not contain any interference bias signals and uses strictly less information than the DD estimator in a two-sided design.

\noindent\emph{Why does TSRN's customer-side normalization introduce bias?} TSRN divides $Q_{11}$ by $\aC\cdot\aL$, implicitly treating the treated-treated cell as a scaled-down version of global treatment. But treated customers face a mixed choice set, both enhanced treatment listings and baseline control listings competing for their attention. This competition is absent under global treatment. TSRN's structure models interference as symmetric across both market sides. By normalizing away $\aC$, TSRN incorrectly assumes the customer-side mixing effect scales linearly, when in fact the nonlinear inventory dynamics may have already absorbed the relevant interference through the listing channel.

In the next sections, we formalize and robustify this intuition. In Section~\ref{sec:theory} we consider a stylized special case of the general model in the small treatment effect regime. We show that provably the DD estimator has smaller bias than the TSRN estimator for most natural market conditions. In Section~\ref{sec:experiments} we perform a broad experimental evaluation of the DD estimator in a variety of market conditions of the general mean field model and showcase that the DD estimator dominates over many other candidate estimators in almost all market settings considered.

\section{Theoretical Analysis}\label{sec:theory}

We now establish our main theoretical results. To obtain closed-form expressions, we specialize to a tractable case that preserves the essential interference dynamics.

\begin{assumption}[Analytical Framework]\label{ass:model}
We assume:
\begin{enumerate}
    \item[(i)] Single customer type and single listing type ($n = m = 1$).
    \item[(ii)] Balanced flow: the arrival rate of customers equals the return rate of occupied listings, $\mu/\tau = 1$.
    \item[(iii)] Normalized control propensity: the matching propensity under control is normalized to $1$.
    \item[(iv)] Small treatment effect: the treatment propensity is $1 + \eps$ for $\eps \geq 0$.
    \item[(v)] Balanced customer randomization: $\aC = \frac{1}{2}$.
    \item[(vi)] Non-trivial listing randomization: $\aL \in (0, 1)$.
\end{enumerate}
\end{assumption}

\noindent The \emph{outside option parameter} $a > 0$ represents the customer's normalized outside option. With the normalization of control propensity to $1$, the choice probability under control and treatment, correspondingly, is:
\begin{align}
\pi_C(s) =~& \frac{s}{a + s}
&
\pi_T(s) =~& \frac{(1+\epsilon)\, s}{a + (1 + \epsilon) s}.
\end{align}
We will analyze the bias of the DD and TSRN estimator in the small treatment effect regime, i.e. $\epsilon \to 0$. Hence, we will begin our analysis by characterizing the true treatment effect for any function of $\epsilon$, as well as the derivative of the effect with respect to $\epsilon$, at $\epsilon=0$.

\subsection{GTE Behavior in Small Treatment Regime}

We first characterize the steady states under global control and treatment (see proofs in Appendix~\ref{app:proofs}).

\begin{lemma}[Control Steady State]\label{lem:control_ss}
Under global control, the steady-state inventory $\hatsC$ satisfies:
\begin{equation}
1 - \hatsC = \frac{\hatsC}{a + \hatsC}.
\end{equation}
The unique positive solution is:
\begin{equation}\label{eq:hatsC}
\hatsC = \frac{-a + \sqrt{a^2 + 4a}}{2} = \frac{-a + \sqrt{a}\sqrt{a+4}}{2}.
\end{equation}
\end{lemma}

\begin{lemma}[Treatment Steady State]\label{lem:treatment_ss}
Under global treatment with propensity $(1+\eps)$, the steady-state inventory $\hatsT$ satisfies:
\begin{equation}
1 - \hatsT = \frac{(1 + \eps)\hatsT}{a + (1 + \eps)\hatsT}.
\end{equation}
The unique positive solution is:
\begin{equation}\label{eq:hatsT}
\hatsT = \frac{-a + \sqrt{a^2 + 4a(1 + \eps)}}{2(1 + \eps)} = \frac{-a + \sqrt{a}\sqrt{a + 4(1 + \eps)}}{2(1 + \eps)}.
\end{equation}
\end{lemma}

\noindent The GTE is the difference in booking rates between global treatment and global control:
\begin{equation}\label{def:TTE}
\TTE(\eps) \coloneqq \frac{(1+\eps)\hatsT}{a + (1+\eps)\hatsT} - \frac{\hatsC}{a + \hatsC}.
\end{equation}
At $\eps = 0$, trivially, we have $\hatsC = \hatsT$, so $\TTE(0) = 0$. The GTE represents the change in steady-state booking rate when treatment makes listings more attractive. Next we characterize the derivative of this attractiveness at $\epsilon=0$.

\begin{lemma}[Derivative of GTE]\label{lem:dTTE}
The derivative of the true treatment effect at $\eps = 0$ is:
\begin{equation}\label{eq:dTTE}
\frac{\dd(\TTE)}{\dd\eps}\bigg|_{\eps=0} = \frac{4a^{3/2}}{\sqrt{a+4}\left(\sqrt{a}\sqrt{a+4} + a\right)^2} > 0 \quad \text{for all } a > 0.
\end{equation}
\end{lemma}

\subsection{Experimental Steady State}

Under the two-sided randomized experiment, let $\sC$ and $\sT$ denote the steady-state inventories of control and treatment listings respectively.
Under the two-sided randomized experiment, the steady-state inventories $\sC$ and $\sT$ satisfy:

\textbf{Control listings} (inflow rate $= 1 - \aL$):
\begin{equation}
(1 - \aL) - s_C^* = \alpha_C \cdot \underbrace{\frac{s_C^*}{a + s_C^* + (1+\varepsilon)s_T^*}}_{\text{treated buyers}} + (1-\alpha_C) \cdot \underbrace{\frac{s_C^*}{a + s_C^* + s_T^*}}_{\text{control buyers}}
\end{equation}

\textbf{Treatment listings} (inflow rate $= \aL$):
\begin{equation}
\aL - s_T^* = \alpha_C \cdot \underbrace{\frac{(1+\varepsilon)s_T^*}{a + s_C^* + (1+\varepsilon)s_T^*}}_{\text{treated buyers}} + (1-\alpha_C) \cdot \underbrace{\frac{s_T^*}{a + s_C^* + s_T^*}}_{\text{control buyers}}
\end{equation}
The factor $(1+\eps)$ appears in the numerator of the first term on the RHS of the treatment listings because treated buyers see treatment listings with an enhanced propensity. It will be convenient in subsequent analysis to define the denominators that appear in the choice probabilities:
\begin{align*}
D_1 &\coloneqq a + \sC + (1 + \eps)\sT  \tag{denominator for treatment buyers}\\
D_2 &\coloneqq a + \sC + \sT \quad \tag{denominator for control buyers}
\end{align*}
The steady-state equations simplify to:
\begin{align}
F_1(\varepsilon, s_C, s_T) &\equiv (1 - \aL - s_C) - \aC \cdot\frac{s_C}{D_1} - (1 - \aC)\cdot\frac{s_C}{D_2} = 0 \label{eq:F1}\\
F_2(\varepsilon, s_C, s_T) &\equiv (\aL - s_T) - \aC \cdot\frac{(1+\varepsilon)s_T}{D_1} - (1-\aC)\cdot\frac{s_T}{D_2} = 0 \label{eq:F2}
\end{align}
and the booking rates for each customer-listing cohort combination simplify to:
\begin{align*}
Q_{11} =~& \aC \cdot \frac{(1+\eps)\sT}{D_1}, & 
Q_{10} =~& \aC \cdot \frac{\sC}{D_1}, &
Q_{01} =~& (1-\aC) \cdot \frac{\sT}{D_2}, &
Q_{00} =~& (1-\aC) \cdot \frac{\sC}{D_2}.
\end{align*}
By Lemma~\ref{lem:equiv-form}, the Double Diagonal estimator reduces to:
\begin{equation}
\GTE_{\text{DD}}(\epsilon) = \frac{Q_{11} + Q_{01}}{\aL} - \frac{Q_{10} + Q_{00}}{1 - \aL} =  \frac{\sC}{1 - \aL} - \frac{\sT}{\aL}.
\end{equation}
Moreover, we remind that the TSRN estimator is:
\begin{equation}
\GTE_{\text{TSRN}}(\epsilon) = \frac{Q_{11}}{\aC\,\aL} - \frac{Q_{01} + Q_{10} + Q_{00}}{1 - \aC\,\aL}.
\end{equation}

\noindent We are interested in understanding the bias of these two estimators and comparing them in the small treatment effect regime.
\begin{definition}[Signed Bias]\label{def:bias}
The signed biases are:
\begin{align*}
\BiasDD(\eps) &\coloneqq \GTE_{\text{DD}}(\eps) - \TTE(\eps), &
\BiasTSRN(\eps) &\coloneqq \GTE_{\text{TSRN}}(\eps) - \TTE(\eps).
\end{align*}
\end{definition}
\noindent Our main theoretical results are summarized in the following theorem:
\begin{theorem}[Main Result]\label{thm:main}
Under Assumption~\ref{ass:model}, fix $a>0$ and $\aL\in(0,1)$. Then:
\begin{enumerate}
    \item[(i)] \textbf{DD Underestimation:} $\BiasDD(\epsilon) < 0$ for all sufficiently small $\eps>0$ (DD underestimates the true treatment effect).
    \item[(ii)] \textbf{TSRN Overestimation:} $\BiasTSRN(\epsilon) > 0$ for all sufficiently small $\eps>0$ (TSRN overestimates the true treatment effect).
    \item[(iii)] \textbf{Threshold:} There is a unique $a^*(\aL) > 0$, not depending on $\eps$, with the following property. If $a < a^*(\aL)$, then $|\BiasDD(\eps)| < |\BiasTSRN(\eps)|$ for all sufficiently small $\eps>0$. If $a > a^*(\aL)$, then $|\BiasDD(\eps)| > |\BiasTSRN(\eps)|$ for all sufficiently small $\eps>0$.
    \item[(iv)] \textbf{Explicit Formula:} $a^*(1/2)=\frac43+\frac{8\sqrt7}{21} \approx 2.34$. So for $\aL=\frac12$, DD's bias is smaller whenever the outside option is at most about twice as attractive as platform options.
    \item[(v)] \textbf{Monotonicity:} $\dd a^*/\dd\aL < 0$ for all $\aL \in (0, 1)$.
\end{enumerate}
\end{theorem}

\noindent In the subsequent sections we un-pack each of the results and provide corresponding intuition and discussion.

\subsection{The Sign of the Bias of DD and TSRN}

As we note in Appendix~\ref{app:eps-zero}, when $\epsilon=0$, intuitively, we have that $\sT=\aL \hatsC$ and $\sC=(1-\aL)\hatsC$ and both estimators return a correct zero effect, i.e. $\GTE_{\text{DD}}(0)=\GTE_{\text{TSRN}}(0)=0=\TTE(0)$. Thus, to understand the behavior of the biases in the small treatment effect regime, we will examine the derivative of the bias at $\epsilon=0$.
\begin{definition}[Bias Derivatives]\label{def:bias_deriv}
Define the first-order bias coefficients:
\begin{align*}
\mathcal{B}_{\text{DD}}(a, \aL) &\coloneqq \frac{\dd(\BiasDD)}{\dd\eps}\bigg|_{\eps=0}, &
\mathcal{B}_{\text{TSRN}}(a, \aL) &\coloneqq \frac{\dd(\BiasTSRN)}{\dd\eps}\bigg|_{\eps=0}.
\end{align*}
\end{definition}

We first show that the derivative of the bias of the DD estimator is negative at $\epsilon=0$. Thus for sufficiently small treatment effects, the DD estimator underestimates the effect. The proof proceeds by computing the explicit expression for $\mathcal{B}_{\text{DD}}$ using implicit differentiation of the steady-state system. A key structural observation is that $\mathcal{B}_{\text{DD}}$ is independent of $\aL$: the $\aL$-dependence in both $\frac{1}{1-\aL}\frac{\dd\sC}{\dd\eps}$ and $\frac{1}{\aL}\frac{\dd\sT}{\dd\eps}$ cancels out exactly after substituting the steady-state values from Lemma~\ref{lem:exp_ss_0}. Under the substitution $t = \sqrt{a}$ and $u = \sqrt{t^2 + 4}$, the expression can be written as a ratio of polynomials in $t$ and $u$ with all coefficients of definite sign, enabling a complete sign analysis (see proof in Appendix~\ref{app:dd_under}).

\begin{theorem}[DD Underestimates]\label{thm:dd_under}
For all $a > 0$ and $\aL \in (0, 1)$:
\begin{equation}
\mathcal{B}_{\text{DD}}(a, \aL) < 0.
\end{equation}
Moreover, $\mathcal{B}_{\text{DD}}$ is independent of $\aL$. Consequently, for small enough $\eps > 0$, the DD estimator underestimates the true treatment effect: 
\[\GTE_{\text{DD}} < \TTE.\]
\end{theorem}

\noindent Using a similar approach, we show that the derivative of the bias of the TSRN estimator is positive at $\epsilon=0$. Unlike $\mathcal{B}_{\text{DD}}$, the TSRN bias derivative does depend on $\aL$. Under the substitution $t = \sqrt{a}$ and $u = \sqrt{t^2 + 4}$, both the numerator and denominator of $\mathcal{B}_{\text{TSRN}}$ are linear in $\aL$, which reduces the bivariate sign analysis to a univariate problem (see proof in Appendix~\ref{app:tsrn_over}).
\begin{theorem}[TSRN Overestimates]\label{thm:tsrn_over}
For all $a > 0$ and $\aL \in (0, 1)$:
\begin{equation}
\mathcal{B}_{\text{TSRN}}(a, \aL) > 0.
\end{equation}
Consequently, for small $\eps > 0$, the TSRN estimator overestimates the true treatment effect: \[\GTE_{\text{TSRN}} > \TTE.\]
\end{theorem}

\subsection{Bias Magnitude Comparison}

Our goal is to understand the difference in the magnitude of the two biases for $\epsilon$ close to zero, i.e.,
\begin{align*}
    |\BiasDD(\eps)| - |\BiasTSRN(\eps)|
\end{align*}
Now that we know the sign of the bias of each estimator for a sufficiently small $\epsilon$, we can reduce the bias comparison by looking at the combined bias function:
\begin{equation*}
\Psi(a, \aL) \coloneqq \mathcal{B}_{\text{DD}}(a, \aL) + \mathcal{B}_{\text{TSRN}}(a, \aL) \tag{Combined Bias Function}
\end{equation*}
Note that by Theorem~\ref{thm:dd_under} and Theorem~\ref{thm:tsrn_over}: 
\begin{align*}
    \Psi(a, \aL) > 0\Leftrightarrow
    -\mathcal{B}_{\text{DD}} < \mathcal{B}_{\text{TSRN}} \Leftrightarrow |\mathcal{B}_{\text{DD}}| < |\mathcal{B}_{\text{TSRN}}|
\end{align*}
Thus the bias of the two estimators is zero at $\epsilon=0$, but the absolute bias of the TSRN grows faster near zero than the bias of the DD estimator. Thus, for a sufficiently small $\epsilon$, we can deduce the desired dominance:
\begin{align*}
    |\BiasDD(\eps)| < |\BiasTSRN(\eps)|
\end{align*}
Thus, it suffices to understand the behavior of the combined bias function $\Psi$. First, we characterize the behavior of $\Psi(\cdot, \aL)$ as the outside option ranges from $0$ to $\infty$ (proof in Appendix~\ref{app:asymp_psi}).  

\begin{lemma}[Boundary Behavior of $\Psi$]\label{lem:asymp}
For fixed $\aL \in (0, 1)$:
\begin{enumerate}
    \item[(i)] As $a \to 0^+$: $\Psi(a, \aL) \to \frac{\aL - 1}{\aL - 2} > 0$.
    \item[(ii)] As $a \to +\infty$: $\Psi(a, \aL) \to 0^-$ (i.e., $\Psi \to 0$ from below).
\end{enumerate}
In particular, $\Psi$ is strictly positive near $a = 0$ and strictly negative for all sufficiently large $a$.
\end{lemma}
\noindent Subsequently, we show that for each $\aL$, the function $\Psi(\cdot, \aL)$ has a unique zero (proof in Appendix~\ref{app:unique_zero}). Since $\Psi$ is positive near $a=0$ (Lemma~\ref{lem:asymp}(i)) and negative for large $a$
(Lemma~\ref{lem:asymp}(ii)), the intermediate value theorem guarantees at least one zero.
Uniqueness is established by writing $\Psi=\mathrm{Num}/\mathrm{Den}$ over the common
denominator $\mathrm{Den}<0$ of the expressions in Theorems~\ref{thm:dd_under}
and~\ref{thm:tsrn_over}, with $\mathrm{Num}=L+M\sqrt{a+4}$. The product of $\mathrm{Num}$ with its
conjugate $L-M\sqrt{a+4}$ is a positive multiple of an explicit quadratic in $a$. Hence the zeros
of $\Psi$ are among the two roots of that quadratic, and each zero is a simple sign change. An
exact evaluation at $\aL=\frac12$ identifies which root is genuine there, and a continuity argument
in $\aL$ extends this to all $\aL\in(0,1)$ (proof in Appendix~\ref{app:unique_zero}).

\begin{theorem}[Uniqueness of Critical Threshold]\label{thm:unique}
For each $\aL \in (0, 1)$, there exists a unique $a^*(\aL) > 0$ such that $\Psi(a^*(\aL), \aL) = 0$. Furthermore:
\begin{enumerate}
    \item[(i)] For $a < a^*(\aL)$: $\Psi > 0$, hence $|\BiasDD| < |\BiasTSRN|$.
    \item[(ii)] For $a > a^*(\aL)$: $\Psi < 0$, hence $|\BiasDD| > |\BiasTSRN|$.
\end{enumerate}
The threshold has the closed-form expression:
\begin{equation}
a^*(\aL) = \frac{4(2\aL^2 - 8\aL + 7) + 4\sqrt{8\aL^2 - 24\aL + 17}}{2(\aL - 2)(\aL - 4)}.
\end{equation}
\end{theorem}

\noindent For the case when $\aL=1/2$, the closed-form from Theorem~\ref{thm:unique} yields an explicit threshold.
\begin{corollary}[Explicit Threshold at $\aL = 1/2$]\label{thm:explicit}
When $\aL = \frac{1}{2}$:
\begin{equation*}
a^* = \frac{4}{3} + \frac{8\sqrt{7}}{21} \approx 2.34.
\end{equation*}
Thus, for $\aL = \frac{1}{2}$ and small $\eps > 0$: DD has lower bias than TSRN if and only if $a < a^* \approx 2.34$.
\end{corollary}

\noindent Moreover, we can also show that the critical threshold $a^*(\aL)$ is a strictly decreasing function of $\aL$. Thus, for the practically relevant regime where the probability of treatment is small, i.e., $\aL \leq 1/2$, the threshold is at least $2.34$. Hence, for any $\aL \leq 1/2$, the DD has smaller bias in the small treatment effect regime than TSRN, as long as the outside option is at most about twice as attractive as the platform option.
\begin{theorem}[Monotonicity of Threshold]\label{thm:mono}
The threshold $a^*(\aL)$ is strictly decreasing in $\aL$:
\begin{equation}
\frac{\dd a^*}{\dd\aL} < 0 \quad \text{for all } \aL \in (0, 1).
\end{equation}
\end{theorem}

\begin{proof}
By the proof of Theorem~\ref{thm:unique}, $a^*(\aL)$ is the larger root of an explicit quadratic
$Q(\cdot,\aL)$ (Appendix~\ref{app:unique_zero}). Differentiating $Q(a^*(\aL),\aL)=0$ gives
\begin{equation}
\frac{\dd a^*}{\dd\aL} = -\frac{\partial Q/\partial\aL}{\partial Q/\partial s}\bigg|_{s=a^*(\aL)}.
\end{equation}
We show in Appendix~\ref{app:monotonicity} that the denominator equals $\sqrt{\Delta(\aL)}>0$ and
the numerator is positive, which implies $\dd a^*/\dd\aL < 0$.

\end{proof}

\section{Experimental Evaluation}\label{sec:experiments}

We evaluate the bias properties of DD, TSRN, CR, LR, and the TSRI-k family across 40 market configurations spanning treatment effect structures, supply-demand dynamics, and market heterogeneity. Apart from all these benchmarks introduced in the prior literature, we also compare against a simpler \emph{Diagonal} estimator as a benchmark:
\begin{equation}\label{eq:diag}
\GTE_{\text{Diag}} = \frac{Q_{11}}{\aC\, \aL} - \frac{Q_{00}}{(1-\aC)(1-\aL)},
\end{equation}
which uses only the on-diagonal cells. This corresponds to a variant of the DD estimator, where the two interference bias correction terms from Equation~\eqref{eqn:interp-dd} are not incorporated. Moreover, we note that the Diagonal estimator is a special case of the TSRI-2 estimator with $\beta = 1/2$. 

\subsection{Experimental Setup}

\paragraph{Market Model.} We instantiate the mean-field model with $n$ customer types and $m$ listing types. Each configuration specifies: (i) matching propensity matrix $P \in [0,1]^{n \times m}$, (ii) treatment matrix $\tilde{P}$, (iii) outside options $\epsilon_i$, (iv) turnover rates $\tau_j$, (v) arrival rate $\mu$, and (vi) type distributions $\phi$, $\rho$. Steady states are computed by simulating dynamics until convergence (gradient norm $< 10^{-4}$). Unless noted otherwise, all experiments use balanced randomization ($\aC = \aL = 0.5$). In the appendix, we present the results for $\aC = \aL = 0.2$, which we see are qualitatively the same.

\paragraph{Metrics.} We report absolute bias 
\begin{align*}
    \text{Absolute Bias} &=|\widehat{\text{GTE}} - \text{GTE}| &
    \text{Relative Bias} &= \frac{|\widehat{\text{GTE}} - \text{GTE}|}{|\text{GTE}|} \times 100\%
\end{align*} 

\paragraph{Scenario Categories.} The 40 scenarios span six categories: (1) \emph{Treatment heterogeneity}: uniform, differential, and opposing effects and  functional form changes (additive vs multiplicative); (2) \emph{Supply-demand regimes}: from extreme supply-constrained to demand-constrained; (3) \emph{Market structure}: sparse, dense, single-type markets, and varying market size and dimensionality; (4) \emph{Distributional heterogeneity}: skewed and power-law distributions; (5) \emph{Structural interventions}: interventions that alter the matching graph or competitive relationships, cannibalization and new matches; (6) \emph{Edge cases}: Parameter asymmetries and outside-option extremes

\subsection{Main Results and Discussion}

\begin{table}[t]
\caption{Summary statistics across 40 scenarios.}
\label{tab:summary}
\centering
\begin{tabular}{lcccccc}
\toprule
& DD & TSRN & CR & LR & TSRI1 & TSRI2\\
\midrule
Lowest $|\text{bias}|$ (count) & \textbf{35/40} & 0/40 & 0/40 & 0/40 & 0/40 & 5/40\\
Median $|\text{rel. bias}|$ (\%) & \textbf{20.5} & 117.1 & 140.7 & 151.2 & 105.0 & 63.6\\
$90^{th}$ percentile  $|\text{rel bias}|$ (\%) & \textbf{87.1} & 794.4 & 1041.4 & 913.1 & 639.2 & 371.6\\
Scenarios with $|\text{rel bias}|<$ 25\% & \textbf{28/40} & 1/40 & 0/40 & 1/40 & 1/40 & 4/40\\
Scenarios with $|\text{rel bias}|>$ 100\% & \textbf{3/40} & 32/40 &  34/40 & 36/40 & 27/40 & 7/40\\
\bottomrule
\end{tabular}
\end{table}

Table~\ref{tab:summary} summarizes performance. DD achieves the lowest bias in the vast majority of scenarios, with median relative bias of 20.5\% versus 117\% for TSRN.

\begin{table}[t]
\caption{Relative bias (\%) by category: median. Bold indicates lowest median.}
\label{tab:category}
\centering
\small
\begin{tabular}{lcccccc}
\toprule
Category & DD & TSRN & CR & LR & TSRI1 & TSRI2\\
\midrule
Treatment heterogeneity & \textbf{14.5} & 113.4  & 136.0 & 150.7 & 100.8 & 63.2\\
Supply-demand regimes & \textbf{94.1} & 3504.4  & 5300.7 & 2674.3 & 2674.3 & 2674.3\\
Market structure & \textbf{20.8} & 119.9  & 151.5 & 151.3 & 107.6 & 64.7\\
Distributional heterogeneity & \textbf{17.4} & 118.7  & 146.0 & 151.6 & 105.2 & 61.7\\
Structural interventions & \textbf{23.0} & 105.2  & 135.6 & 119.9 & 96.7 & 63.2\\
Edge cases & \textbf{39.9} & 93.1  & 98.3 & 187.4 & 107.5 & 63.6\\
\bottomrule
\end{tabular}
\end{table}

\paragraph{Performance by Category.}

 The 40 scenarios span six categories: (1) \emph{Treatment heterogeneity} (S1--S5, S14--S15, S17--S18, S21, S23, S31--S33): uniform, differential, and opposing effects and  functional form changes (additive vs multiplicative); (2) \emph{Supply-demand regimes} (S6--S7, S25, S36--S37): from extreme supply-constrained ($\mu = 10$, $\tau = 0.02$) to demand-constrained ($\mu = 0.01$, $\tau = 5.0$); (3) \emph{Market structure} (S8--S11, S28, S34--S35): sparse, dense, single-type markets, and varying market size and dimensionality; (4) \emph{Distributional heterogeneity} (S12--S15, S29--S30): skewed and power-law distributions; (5) \emph{Structural interventions} (S3, S21, S24, S26--S27, S39--S40): interventions that alter the matching graph or competitive relationships, cannibalization and new matches; (6) \emph{Edge cases} (S16, S19, S20, S22, S38): Parameter asymmetries and outside-option extremes

Table~\ref{tab:category} disaggregates results. DD achieves the lowest median bias in all categories, with particularly strong performance in treatment heterogeneity (14.5\%), distributional heterogeneity (17.4\%), and market structure (20.8\%). TSRN, CR, and LR exceed 100\% median bias in most categories.

\begin{figure}[t]
\centering
\includegraphics[width=0.85\textwidth]{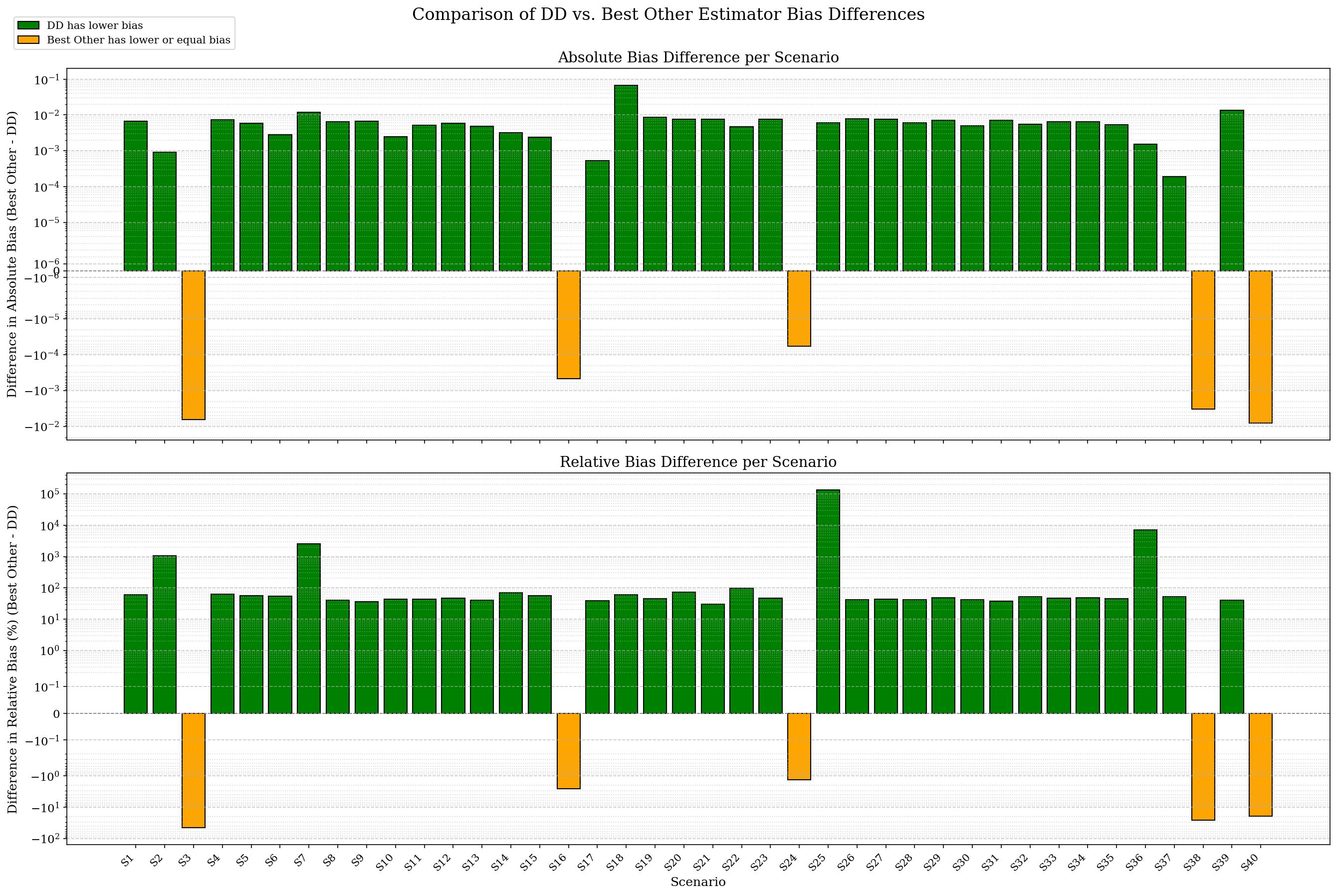}
\caption{Difference in absolute bias (top) and relative bias (bottom) between DD and lowest-bias other estimator per scenario across 40 scenarios. DD attains lowest bias in 35/40 scenarios. Green bars indicate DD has lower bias.}
\label{fig:scenarios}
\end{figure}

Figure~\ref{fig:scenarios} visualizes bias across all scenarios. DD consistently achieves lower bias than alternatives in most scenarios. 

\paragraph{When DD Excels.} DD achieves very low relative bias with uniform treatment effects (S1: 0.1\%) and heterogeneous outside options (S20: 0.3\%). 

\paragraph{When DD Struggles.} With opposing treatment effects (S3), DD exhibits 44.5\% bias while TSRN achieves 34.5\%. When outside options dominate (S38), DD underestimates by 42.3\% versus TSRN's 22\% overestimation.

\begin{figure}[t]
\centering
\includegraphics[width=0.85\textwidth]{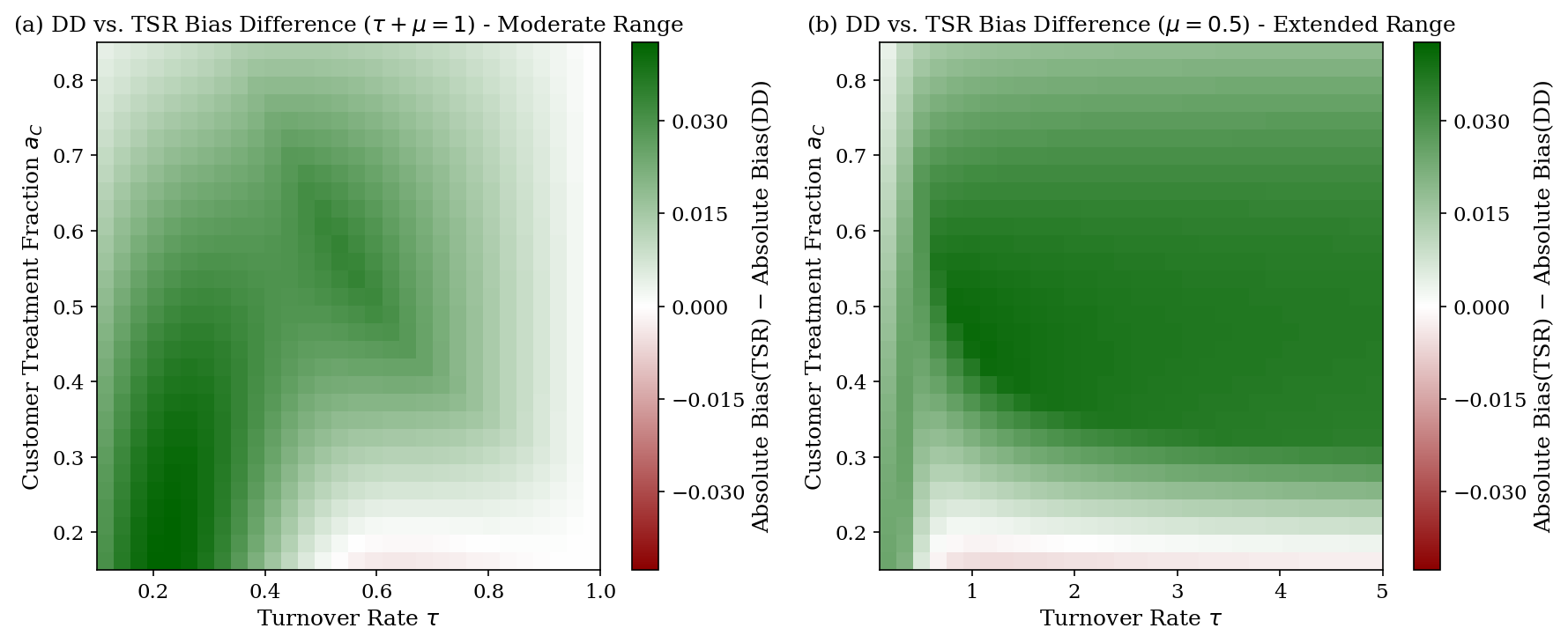}
\caption{$\text{ Signed difference |Bias(TSRN)| - |Bias(DD)| }$ versus turnover rate $\tau$. Low $\tau$: supply-constrained; high $\tau$: demand-constrained. DD achieves lower bias than TSRN across most regimes and allocation fractions.}
\label{fig:turnover}
\end{figure}

\paragraph{Varying Supply-Demand Balance.} Figure~\ref{fig:turnover} traces bias as turnover rate $\tau$ varies from supply-constrained to demand-constrained conditions. DD consistently achieves lower bias than TSRN across all $\tau$ and all tested allocation fractions. The DD-TSRN gap is largest when $\tau$ is small.

\begin{figure}[t]
\centering
\includegraphics[width=0.55\textwidth]{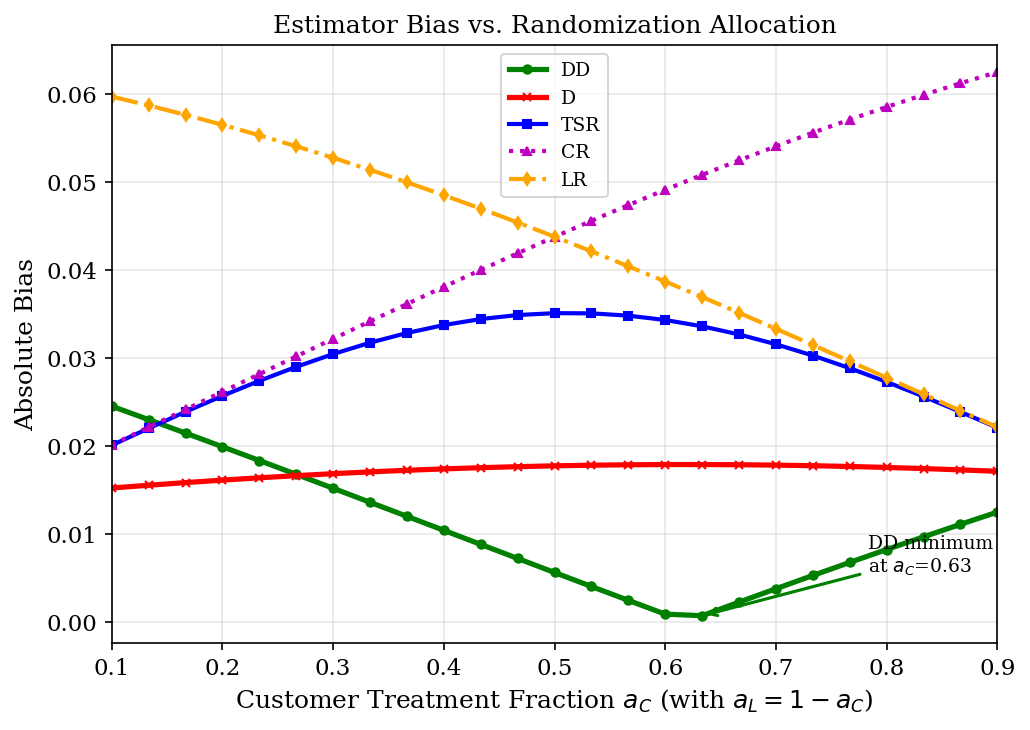}
\caption{Bias versus customer treatment fraction $\aC$ with $\aL = 1 - \aC$. DD achieves minimum near balanced allocation ($\aC \approx 0.5$).}
\label{fig:fraction}
\end{figure}

\paragraph{Varying Randomization Proportions.} Figure~\ref{fig:fraction} shows bias versus $\aC$. DD achieves minimum bias near $\aL = \aC = 0.5$, increasing under imbalance. The one-sided estimators exhibit asymmetries: LR struggles at low $\aC$; CR struggles at high $\aC$. TSRN maintains flat bias but exceeds DD in most conditions.

\begin{figure}[t]
\centering
\includegraphics[width=0.7\textwidth]{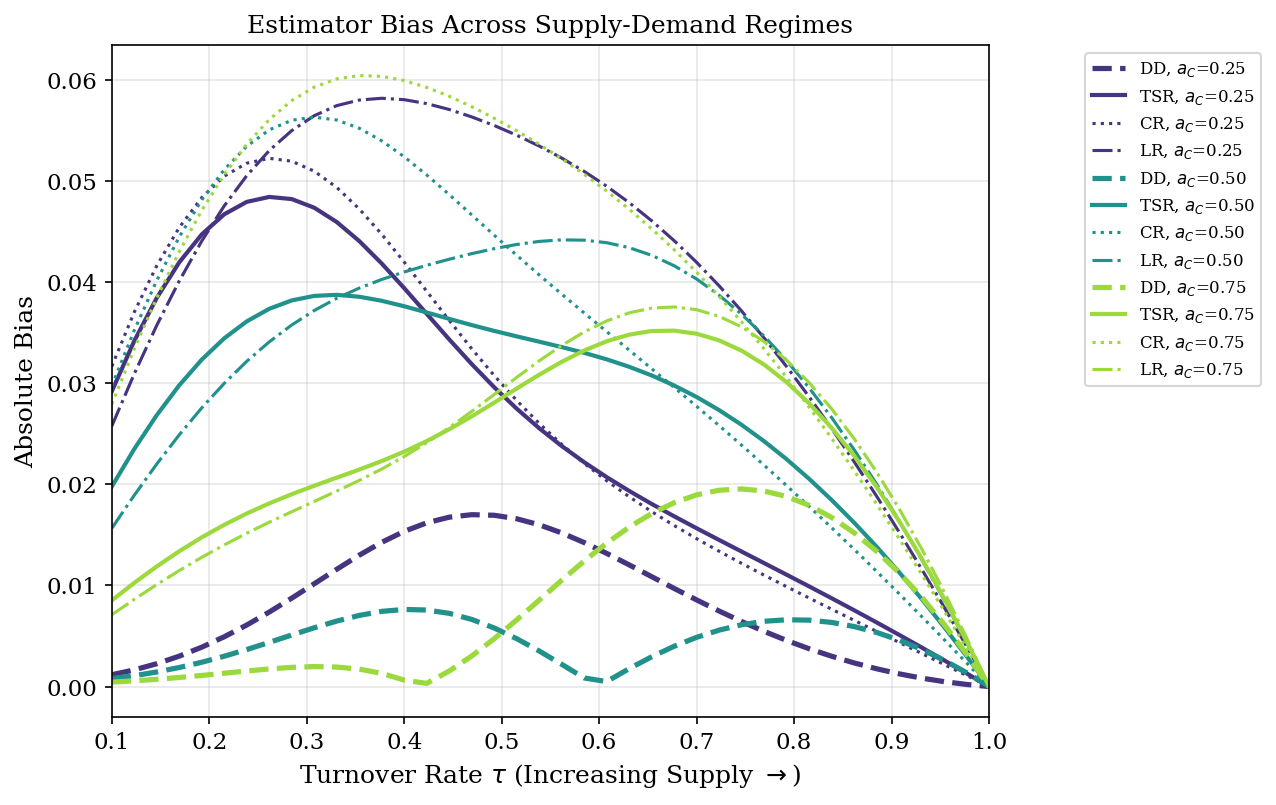}
\caption{All four estimators across market regimes. The ordering DD $<$ TSRN $<$ CR $\approx$ LR holds consistently.}
\label{fig:four}
\end{figure}

\paragraph{Four-Estimator Comparison.} Figure~\ref{fig:four} compares all estimators across the turnover range. The ordering DD $<$ TSRN $<$ CR $\approx$ LR holds consistently. Two main findings emerge from our experiments:

\emph{(i) DD achieves lowest bias across most regimes.} DD achieves lower bias in the vast majority of configurations. Parameter sweeps confirm this is not scenario-specific: DD dominates for every supply level and allocation fraction tested.

\emph{(ii) Opposite bias directions.} TSRN, CR, and LR consistently overestimate when treatment effects are positive, while DD underestimates. This confirms our theoretical finding that the two estimators exhibit biases in opposite directions.

\subsection{Limitations}

Our theoretical results apply to a simplified homogeneous market model with specific parameter restrictions. While our numerical experiments suggest the qualitative findings extend broadly, formal theoretical analysis of heterogeneous markets remains an open direction. Moreover, we have only focused on bias and ignored any variance consideration. Hence, our analysis is most useful in large sample regimes (relevant for online platforms) where variance is of lower order importance. However, a complete comparison in the small sample regime also needs to account for variance, in order to capture all potentially leading terms of the mean squared error. Systematic variance analysis is beyond our current scope. Finally, our model assumes that the intervention affects only matching propensities, not arrival rates or occupancy times. Extensions to these intervention types represent important future work.

\section{Conclusion}
\label{sec:conclusion}

We have introduced the Double Diagonal (DD) estimator for treatment effect estimation in two-sided marketplace experiments and provided the first theoretical comparison between DD and the standard naive TSR (TSRN) estimator under a mean-field market model.

Our stylized theoretical analysis establishes two key results: (i) DD's bias is strictly smaller than TSRN's below a critical threshold of the outside-option parameter in the small treatment effect regime; in particular, DD's bias is smaller whenever the outside option is at most about twice as attractive as platform options; and (ii) the two estimators exhibit biases in opposite directions, with DD underestimating and TSRN overestimating the true effect.

Our empirical evaluation across 40 diverse market configurations corroborates these theoretical findings. Across a wide range of configurations, the DD estimator demonstrates the lowest bias, establishing it as a robust and often superior choice for reliable experimentation in two-sided marketplaces.

\begin{acks}

	The authors would like to thank Ramesh Johari for early feedback and discussions.

	Vasilis Syrgkanis is supported by the National Science Foundation Award under Grant
	No.:~\grantnum{IIS-2337916}{IIS-2337916}.

\end{acks}

\bibliographystyle{ACM-Reference-Format}
\bibliography{references}

\newpage 
\appendix
\section{Proofs of Main Results}
\label{app:proofs}

Throughout this appendix, we work under Assumption~\ref{ass:model}: $n = m = 1$, $\mu/\tau = 1$, control propensity normalized to $1$, treatment propensity $1 + \eps$ for $\eps \geq 0$, $\aC = \frac{1}{2}$, $\aL \in (0,1)$, and outside option parameter $a > 0$.

\subsection{Proof of Lemma~\ref{lem:control_ss}}

\begin{proof}
Under global control, the matching propensity is $P = 1$ and the market has unit mass of listings ($\rho = 1$). By Assumption~1 ($\tau = 1$) and Assumption~\ref{ass:model}(ii) ($\mu = 1$), the steady-state condition~\eqref{eq:steady_state} requires inflow to equal outflow:
\begin{equation}
\underbrace{(1 - \hatsC) \cdot 1}_{\text{inflow}} = \underbrace{1 \cdot \frac{\hatsC}{a + \hatsC}}_{\text{outflow}},
\end{equation}
where the outflow is the booking rate under the choice probability~\eqref{eq:choice_prob} with $P = 1$ and $s = \hatsC$. Cross-multiplying yields $(1 - \hatsC)(a + \hatsC) = \hatsC$. Expanding:
\begin{equation}
a + \hatsC - a\hatsC - \hatsC^2 = \hatsC \implies \hatsC^2 + a\hatsC - a = 0.
\end{equation}
The quadratic formula gives $\hatsC = \frac{-a \pm \sqrt{a^2 + 4a}}{2}$. Since $\hatsC > 0$, we take the positive root and factor $\sqrt{a^2 + 4a} = \sqrt{a}\sqrt{a+4}$ (valid for $a > 0$) to obtain~\eqref{eq:hatsC}.
\end{proof}

\subsection{Proof of Lemma~\ref{lem:treatment_ss}}

\begin{proof}
Under global treatment, the matching propensity is $\tilde{P} = 1 + \eps$ (Assumption~\ref{ass:model}(iv)). With $\rho = 1$, $\tau = 1$, $\mu = 1$, the steady-state condition becomes:
\begin{equation}
(1 - \hatsT) = \frac{(1 + \eps)\hatsT}{a + (1 + \eps)\hatsT}.
\end{equation}
Substituting $w = (1 + \eps)\hatsT$ gives $1 - \frac{w}{1 + \eps} = \frac{w}{a + w}$. Multiplying through by $(1+\eps)(a+w)$ and simplifying:
\begin{equation}
(1+\eps)a + (1+\eps)w - wa - w^2 = w(1+\eps),
\end{equation}
which reduces to $w^2 + aw - a(1 + \eps) = 0$. The positive root is $w = \frac{-a + \sqrt{a^2 + 4a(1 + \eps)}}{2}$, and $\hatsT = w/(1+\eps)$ yields~\eqref{eq:hatsT}. At $\eps = 0$, this reduces to $\hatsT|_{\eps=0} = \frac{-a + \sqrt{a}\sqrt{a+4}}{2} = \hatsC$ by Lemma~\ref{lem:control_ss}.
\end{proof}

\subsection{Proof of Lemma~\ref{lem:dTTE}}

\begin{proof}
Define $w(\eps) \coloneqq (1+\eps)\hatsT(\eps) = \frac{-a + \sqrt{a^2 + 4a(1+\eps)}}{2}$, using the expression from the proof of Lemma~\ref{lem:treatment_ss}. Then $Q_T(\eps) = \frac{w}{a+w}$, and since $\hatsC$ (Lemma~\ref{lem:control_ss}) is independent of $\eps$, we have $\frac{\dd(\TTE)}{\dd\eps} = \frac{\dd Q_T}{\dd\eps}$ by~\eqref{def:TTE}. Differentiating:
\begin{equation}
\frac{\dd w}{\dd\eps} = \frac{a}{\sqrt{a^2 + 4a(1+\eps)}}, \qquad \frac{\dd Q_T}{\dd\eps} = \frac{a}{(a+w)^2} \cdot \frac{\dd w}{\dd\eps}.
\end{equation}
At $\eps = 0$: $w|_{\eps=0} = \hatsC = \frac{-a + \sqrt{a}\sqrt{a+4}}{2}$ by Lemma~\ref{lem:control_ss}, so
\begin{equation}
\frac{\dd w}{\dd\eps}\bigg|_{\eps=0} = \frac{a}{\sqrt{a}\sqrt{a+4}} = \frac{\sqrt{a}}{\sqrt{a+4}}, \qquad a + w\big|_{\eps=0} = \frac{a + \sqrt{a}\sqrt{a+4}}{2}.
\end{equation}
Substituting:
\begin{equation}
\frac{\dd(\TTE)}{\dd\eps}\bigg|_{\eps=0} = \frac{a}{\left(\frac{a + \sqrt{a}\sqrt{a+4}}{2}\right)^2} \cdot \frac{\sqrt{a}}{\sqrt{a+4}} = \frac{4a^{3/2}}{\sqrt{a+4}\left(\sqrt{a}\sqrt{a+4} + a\right)^2} > 0,
\end{equation}
since $a > 0$ ensures every factor is strictly positive.
\end{proof}

\subsection{Characterization of Experimental Steady State at \texorpdfstring{$\epsilon=0$}{epsilon = 0}}\label{app:eps-zero}
\noindent The next lemma derives the very intuitive statement that at $\epsilon=0$, the steady state booking rate for control and treatment listings is simply an appropriately scaled version of the global control steady state $\hatsC$:
\begin{lemma}[Experimental Steady State at $\eps = 0$]\label{lem:exp_ss_0}
At $\eps = 0$, define:
\begin{align*}
\phi &\coloneqq \frac{D}{N}, &
\begin{aligned}
    D &\coloneqq \frac{a + \sqrt{a^2 + 4a}}{2} = \frac{a + \sqrt{a}\sqrt{a+4}}{2}, \\
N &\coloneqq D + 1 = \frac{a + 2 + \sqrt{a}\sqrt{a+4}}{2},
\end{aligned}
\end{align*}
Then the experimental steady states at $\eps = 0$ are:
\begin{equation}\label{eq:ss_at_0}
\sC\big|_{\eps=0} = (1 - \aL)\phi, \qquad \sT\big|_{\eps=0} = \aL\phi.
\end{equation}
Moreover, $\phi=\hatsC$, hence, $\sC  = (1 - \aL)\hatsC$ and 
    $\sT = \aL \hatsC$ and $D_1\mid_{\epsilon=0} = D_2\mid_{\epsilon=0} = a + \phi$.
\end{lemma}

\begin{proof}
At $\varepsilon = 0$, $D_1 = D_2 \equiv D = a + s_C^* + s_T^*$. The steady-state equations simplify:
\begin{align}
F_1: \quad (1 - \aL) - s_C^* &= \frac{s_C^*}{D}, \\
F_2: \quad \aL - s_T^* &= \frac{s_T^*}{D}.
\end{align}

Adding these equations:
\begin{equation}
1 - (s_C^* + s_T^*) = \frac{s_C^* + s_T^*}{D}.
\end{equation}

Let $S = s_C^* + s_T^*$ denote the total inventory. Since $D = a + S$:
\begin{equation}
(1 - S) = \frac{S}{a + S}.
\end{equation}

Multiplying both sides by $(a + S)$:
\begin{equation}
(1-S)(a+S) = S \implies a + S - aS - S^2 = S \implies S^2 + aS - a = 0.
\end{equation}

By the quadratic formula:
\begin{equation}
S = \frac{-a + \sqrt{a^2 + 4a}}{2}.
\end{equation}

Comparing with Lemma~\ref{lem:control_ss}, we observe that $S$ and $\hatsC$ satisfy the identical quadratic equation $x^2 + ax - a = 0$, and both are the positive root. Therefore:
\begin{equation}
S = \hatsC.
\end{equation}

From the first equation:
\begin{equation}
(1 - \aL) - \sC = \frac{\sC}{D} \implies (1 - \aL) = \sC\left(1 + \frac{1}{D}\right) = \sC \cdot \frac{D + 1}{D}.
\end{equation}

With $D = a + S$ and $N = D + 1 = a + S + 1$:
\begin{equation}
\sC = \frac{(1 - \aL)D}{N}.
\end{equation}

Similarly, from the second equation: $\sT = \frac{\aL D}{N}$.

We have:
\begin{equation}
\sC + \sT = \frac{D}{N} = \phi.
\end{equation}

But $S = \sC + \sT$, so $\phi = S = \hatsC$.

Substituting:
\begin{equation}
\sC = (1 - \aL)\hatsC, \qquad \sT = \aL\hatsC. \qedhere
\end{equation}
\end{proof}

\subsection{Implicit Differentiation and Bias Derivatives}

The steady-state system $F_1 = F_2 = 0$ (equations~\eqref{eq:F1}--\eqref{eq:F2} with $\aC = \frac{1}{2}$) implicitly defines $\sC(\eps)$ and $\sT(\eps)$. The map $F$ is continuously differentiable near the $\eps=0$ steady state of Lemma~\ref{lem:exp_ss_0}, and its Jacobian $J$ there is nonsingular (Lemma~\ref{lem:jacobian}). The implicit function theorem therefore gives a continuously differentiable solution $(\sC(\eps),\sT(\eps))$ of $F=0$ for $\eps$ near $0$. By uniqueness of the steady state, this solution is the experimental steady state, and
\begin{equation}\label{eq:IFT}
\begin{pmatrix} \frac{\dd\sC}{\dd\eps} \\[4pt] \frac{\dd\sT}{\dd\eps} \end{pmatrix}\bigg|_{\eps=0} = -J^{-1} \begin{pmatrix} B_1 \\ B_2 \end{pmatrix},
\end{equation}
where $J$ and $(B_1, B_2)$ are evaluated at $\eps = 0$ and the steady state from Lemma~\ref{lem:exp_ss_0}.

\begin{lemma}\label{lem:jacobian}
The Jacobian of $F=(F_1, F_2)$ with respect to $(\sC, \sT)$ at $\eps = 0$ is:
\begin{equation}
J = \begin{pmatrix} A_{11} & A_{12} \\ A_{21} & A_{22} \end{pmatrix}, \qquad
\begin{aligned}
A_{11} &= -1 - \frac{a + \sT}{D^2}, \quad & A_{12} &= \frac{\sC}{D^2}, \\
A_{21} &= \frac{\sT}{D^2}, \quad & A_{22} &= -1 - \frac{a + \sC}{D^2},
\end{aligned}
\end{equation}
with $D = a + \sC + \sT$. Moreover,
\[
\det J = 1+\frac{a+D}{D^2}+\frac{a}{D^3}>0,
\]
so $J$ is nonsingular.
\end{lemma}

\begin{proof}
At $\eps = 0$, $D_1 = D_2 = D$ by Lemma~\ref{lem:exp_ss_0}, so $F_1 = (1-\aL) - \sC - \sC/D$ and $F_2 = \aL - \sT - \sT/D$. For $A_{11}$: by the quotient rule with $\frac{\partial D}{\partial \sC} = 1$,
\begin{equation}
\frac{\partial}{\partial \sC}\left(\frac{\sC}{D}\right) = \frac{D - \sC}{D^2} = \frac{a + \sT}{D^2},
\end{equation}
so $A_{11} = -1 - (a + \sT)/D^2$. For $A_{12}$: $\frac{\partial}{\partial \sT}(\sC/D) = -\sC/D^2$, so $A_{12} = \sC/D^2$ (the minus sign from $F_1$ cancels). The entries $A_{21}$ and $A_{22}$ follow by the same argument applied to $F_2 = \aL - \sT - \sT/D$.
For the determinant, write $A_{11}=-(1+x)$ and $A_{22}=-(1+y)$ with $x=(a+\sT)/D^2$ and $y=(a+\sC)/D^2$. Then
\[
\det J=(1+x)(1+y)-\frac{\sC\sT}{D^4}=1+x+y+\frac{(a+\sT)(a+\sC)-\sC\sT}{D^4}.
\]
Since $x+y=(a+D)/D^2$ and $(a+\sT)(a+\sC)-\sC\sT=a(a+\sC+\sT)=aD$, this equals $1+(a+D)/D^2+a/D^3$, which is positive because $a,D>0$.

\end{proof}

\begin{lemma}\label{lem:forcing}
The partial derivatives of $F=(F_1, F_2)$ with respect to $\eps$ at $\eps = 0$ are:
\begin{align}
B_1 \coloneqq \frac{\partial F_1}{\partial \eps}\bigg|_{\eps=0} = \frac{\sC\sT}{2D^2}, \qquad
B_2 \coloneqq \frac{\partial F_2}{\partial \eps}\bigg|_{\eps=0} = -\frac{\sT(a + \sC)}{2D^2}.
\end{align}
\end{lemma}

\begin{proof}
In $F_1$ (equation~\eqref{eq:F1} with $\aC = 1/2$), the only $\eps$-dependent term is $\frac{1}{2}\sC/D_1$ via $D_1 = a + \sC + (1+\eps)\sT$. Since $\frac{\partial D_1}{\partial\eps} = \sT$:
\begin{equation}
\frac{\partial F_1}{\partial \eps} = -\frac{1}{2}\sC\left(-\frac{\sT}{D_1^2}\right) = \frac{\sC\sT}{2D_1^2}.
\end{equation}
At $\eps = 0$: $D_1 = D$ by Lemma~\ref{lem:exp_ss_0}, giving $B_1 = \sC\sT/(2D^2)$.

In $F_2$, the $\eps$-dependent term is $\frac{1}{2}(1+\eps)\sT/D_1$. Differentiating:
\begin{equation}
\frac{\partial}{\partial\eps}\left(\frac{(1+\eps)\sT}{D_1}\right) = \frac{\sT D_1 - (1+\eps){\sT}^2}{D_1^2} = \frac{\sT(a + \sC)}{D_1^2},
\end{equation}
using $D_1 - (1+\eps)\sT = a + \sC$. At $\eps = 0$: $B_2 = -\sT(a + \sC)/(2D^2)$.
\end{proof}

\begin{corollary}[Steady-State Derivatives]\label{cor:cramer}
By the implicit function theorem~\eqref{eq:IFT} and the standard $2\times 2$ inverse formula:
\begin{align}
\frac{\dd\sC}{\dd\eps}\bigg|_{\eps=0} &= \frac{-B_1 A_{22} + B_2 A_{12}}{\det(J)}, \label{eq:dsC}\\
\frac{\dd\sT}{\dd\eps}\bigg|_{\eps=0} &= \frac{-B_2 A_{11} + B_1 A_{21}}{\det(J)}, \label{eq:dsT}
\end{align}
where $A_{ij}$ are from Lemma~\ref{lem:jacobian}, $B_i$ from Lemma~\ref{lem:forcing}, and $\det(J) = A_{11}A_{22} - A_{12}A_{21}$.
\end{corollary}

\subsection{Proof of Theorem~\ref{thm:dd_under}}\label{app:dd_under}

\begin{proof}
By Definition~\ref{def:bias_deriv} and Lemma~\ref{lem:equiv-form} (which gives $\GTE_{\text{DD}} = \frac{\sC}{1-\aL} - \frac{\sT}{\aL}$ in the single-type case):
\begin{equation}\label{eq:BDD-expr}
\mathcal{B}_{\text{DD}} = \frac{1}{1-\aL}\frac{\dd\sC}{\dd\eps}\bigg|_{\eps=0} - \frac{1}{\aL}\frac{\dd\sT}{\dd\eps}\bigg|_{\eps=0} - \frac{\dd(\TTE)}{\dd\eps}\bigg|_{\eps=0}.
\end{equation}
Substituting the steady-state derivatives from Corollary~\ref{cor:cramer} (equations~\eqref{eq:dsC}--\eqref{eq:dsT}), with Jacobian entries from Lemma~\ref{lem:jacobian} and forcing terms from Lemma~\ref{lem:forcing}, and evaluating at the steady state from Lemma~\ref{lem:exp_ss_0}---namely $\sC = (1-\aL)\hatsC$, $\sT = \aL\hatsC$, $D = a + \hatsC$ where $\hatsC$ is given by Lemma~\ref{lem:control_ss}---the $\aL$-dependence in the Jacobian entries and forcing terms takes the form:
\begin{align}
A_{11} &= -1 - \frac{a + \aL\hatsC}{(a+\hatsC)^2}, \qquad & A_{12} &= \frac{(1-\aL)\hatsC}{(a+\hatsC)^2}, \\
A_{21} &= \frac{\aL\hatsC}{(a+\hatsC)^2}, \qquad & A_{22} &= -1 - \frac{a + (1-\aL)\hatsC}{(a+\hatsC)^2}, \\
B_1 &= \frac{\aL(1-\aL)\hatsC^2}{2(a+\hatsC)^2}, \qquad & B_2 &= -\frac{\aL\hatsC(a + (1-\aL)\hatsC)}{2(a+\hatsC)^2}.
\end{align}
The factors of $\aL$ and $(1-\aL)$ in $B_1$, $B_2$, $A_{12}$, and $A_{21}$ combine with the $\frac{1}{1-\aL}$ and $\frac{1}{\aL}$ prefactors in~\eqref{eq:BDD-expr} so that every $\aL$-dependent term cancels exactly after algebraic simplification. (This cancellation is verified symbolically.) The GTE derivative from Lemma~\ref{lem:dTTE} depends only on $a$, confirming that $\mathcal{B}_{\text{DD}}$ is independent of $\aL$.

Under the substitution $t = \sqrt{a}$ and $u = \sqrt{t^2 + 4}$---so that $\hatsC = \frac{-t^2 + tu}{2}$ and $a + \hatsC = \frac{t(t+u)}{2}$ by Lemma~\ref{lem:control_ss}, the expression reduces to:
\begin{equation}
\mathcal{B}_{\text{DD}} = \frac{N(t)}{D(t)},
\end{equation}
where
\begin{align}
N(t) &= -t^6 - 5t^4 - 4t^2 + u\bigl(-t^5 - 3t^3\bigr), \\
D(t) &= 2t^8 + 20t^6 + 66t^4 + 76t^2 + 16 + u\bigl(2t^7 + 16t^5 + 38t^3 + 24t\bigr).
\end{align}

\textbf{Numerator $N(t) < 0$.} Factoring:
\begin{equation}
N(t) = -t^2 \bigl[t^4 + 5t^2 + 4 + u(t^3 + 3t)\bigr].
\end{equation}
For $t > 0$: $t^2 > 0$, and every term inside the bracket is strictly positive since $t > 0$ and $u = \sqrt{t^2 + 4} > 0$. Therefore $N(t) < 0$.

\textbf{Denominator $D(t) > 0$.} Every coefficient is positive, and $t > 0$, $u > 0$, so $D(t) > 0$.

Therefore $\mathcal{B}_{\text{DD}} = N(t)/D(t) < 0$ for all $a > 0$.
\end{proof}

\subsection{Proof of Theorem~\ref{thm:tsrn_over}: TSRN Overestimates}
\label{app:tsrn_over}

\begin{proof}
We compute $\mathcal{B}_{\text{TSRN}}$ (Definition~\ref{def:bias_deriv}) by differentiating the TSRN estimator~\eqref{eq:TSRN} with $\aC = \frac{1}{2}$. Since each booking rate $Q_{dz}$ depends on $\eps$ both directly (through the factor $(1+\eps)$ in $Q_{11}$ and through $D_1$) and indirectly (through $\sC(\eps)$ and $\sT(\eps)$), we apply the chain rule, writing $\dot{s}_C \coloneqq \dd\sC/\dd\eps|_{\eps=0}$ and $\dot{s}_T \coloneqq \dd\sT/\dd\eps|_{\eps=0}$ from Corollary~\ref{cor:cramer}:
\begin{align}
\dot{Q}_{11} &= \tfrac{1}{2}\cdot\frac{\sT(a+\sC) + (a+\sC)\dot{s}_T - \sT\dot{s}_C}{D^2}, \label{eq:dQ11}\\
\dot{Q}_{10} &= \tfrac{1}{2}\cdot\frac{(a+\sT)\dot{s}_C - \sC(\sT + \dot{s}_T)}{D^2}, \label{eq:dQ10}\\
\dot{Q}_{01} &= \tfrac{1}{2}\cdot\frac{(a+\sC)\dot{s}_T - \sT\dot{s}_C}{D^2}, \label{eq:dQ01}\\
\dot{Q}_{00} &= \tfrac{1}{2}\cdot\frac{(a+\sT)\dot{s}_C - \sC\dot{s}_T}{D^2}. \label{eq:dQ00}
\end{align}
In~\eqref{eq:dQ11}, the term $\sT(a+\sC)$ arises from the direct differentiation of the factor $(1+\eps)$ at $\eps = 0$; the remaining terms come from differentiating $\sT/D_1$ via the quotient rule (noting $\partial D_1/\partial\eps = \sT$ at $\eps = 0$) and the chain rule through $\sC(\eps), \sT(\eps)$. In~\eqref{eq:dQ10}, the $\sC \cdot \sT$ contribution in the numerator similarly arises from the direct $\eps$-dependence of $D_1$. Expressions~\eqref{eq:dQ01}--\eqref{eq:dQ00} involve only the indirect dependence through $\sC(\eps), \sT(\eps)$, since $D_2 = a + \sC + \sT$ does not depend on $\eps$ directly.

The first-order bias coefficient is then:
\begin{equation}
\mathcal{B}_{\text{TSRN}} = \frac{\dot{Q}_{11}}{\frac{1}{2}\aL} - \frac{\dot{Q}_{01}+\dot{Q}_{10}+\dot{Q}_{00}}{1-\frac{1}{2}\aL} - \frac{\dd(\TTE)}{\dd\eps}\bigg|_{\eps=0},
\end{equation}
where $\frac{\dd(\TTE)}{\dd\eps}\big|_{\eps=0}$ is from Lemma~\ref{lem:dTTE}. We substitute $\sC = (1-\aL)\hatsC$, $\sT = \aL\hatsC$, $D = a + \hatsC$ from Lemma~\ref{lem:exp_ss_0}, $\hatsC$ from Lemma~\ref{lem:control_ss}, and $\dot{s}_C$, $\dot{s}_T$ from Corollary~\ref{cor:cramer}.

Under the substitution $t = \sqrt{a}$ and $u = \sqrt{t^2+4}$, symbolic computation yields $\mathcal{B}_{\text{TSRN}} = N(t,\aL)/D(t,\aL)$, where both numerator and denominator are of the form $P_0(t) + P_1(t)\cdot\aL + [Q_0(t)+Q_1(t)\cdot\aL]\cdot u$, that is, \emph{linear} in $\aL$. This linearity reduces the bivariate sign analysis to two univariate problems.

\paragraph{Sign of the denominator.} Writing $D = r_0(t) + r_1(t)\aL + [s_0(t)+s_1(t)\aL]\cdot u$, symbolic computation reveals $r_0 = -2r_1$ and $s_0 = -2s_1$, yielding:
\begin{equation}
D(t,\aL) = (\aL-2)\bigl[r_1(t) + s_1(t)\cdot u\bigr].
\end{equation}
The polynomials $r_1(t)$ and $s_1(t)$ have all positive coefficients (see Appendix~\ref{explicit_poly}), so the bracket is strictly positive for $t>0$, $u>0$. Since $\aL\in(0,1)$ implies $\aL-2<0$, we have $D(t,\aL)<0$.

\paragraph{Sign of the numerator.} Writing $N = p_0(t) + p_1(t)\aL + [q_0(t)+q_1(t)\aL]\cdot u$, we analyze the slope and a boundary value.

\emph{Slope in $\aL$:} The coefficient $p_1(t) + q_1(t)\cdot u$ is strictly positive for all $t>0$, since $p_1$ and $q_1$ have all positive coefficients (Appendix~\ref{explicit_poly}). Hence $N$ is strictly increasing in $\aL$ on $(0,1)$.

\emph{Value at $\aL=1$:} The sums $p_1+p_0$ and $q_1+q_0$ have all non-positive coefficients (Appendix~\ref{explicit_poly}), with at least one strictly negative coefficient of positive degree. Therefore $N|_{\aL=1} = (p_1+p_0) + (q_1+q_0)\cdot u < 0$ for $t>0$.

Since $N$ is increasing in $\aL$ and strictly negative at $\aL=1$, we conclude $N(t,\aL) < 0$ for all $\aL\in(0,1)$ and $t>0$.

\paragraph{Conclusion.}
$N < 0$ and $D < 0$ imply $\mathcal{B}_{\text{TSRN}} = N/D > 0$. For small $\eps>0$: $\BiasTSRN(\eps) \approx \eps\cdot\mathcal{B}_{\text{TSRN}} > 0$, confirming that TSRN overestimates the true treatment effect.
\end{proof}

\subsection{Proof of Lemma~\ref{lem:asymp}: Boundary Behaviour of \texorpdfstring{$\Psi$}{Psi}}
\label{app:asymp_psi}

\begin{proof}
Write $t=\sqrt a$ and $u=\sqrt{a+4}$. By the proofs of Theorems~\ref{thm:dd_under}
and~\ref{thm:tsrn_over}, $\Psi=N(t)/D(t)+N(t,\aL)/D(t,\aL)$, where $D(t)=d_0(t)+d_1(t)\,u$ and the
remaining polynomials are listed in Appendix~\ref{explicit_poly}.

\paragraph{Part (i): $a\to 0^+$.}
As $a\to0^+$ we have $t\to0$ and $u\to2$. Each $u$-coefficient ($t^5+3t^3$, $d_1$, $q_1$, $q_0$,
$s_1$) has no constant term, so it vanishes at $t=0$. The $u$-free parts at $t=0$ are
$N(0)=0$, $d_0(0)=16$, $\aL p_1(0)+p_0(0)=8\aL-8$ and $(\aL-2)r_1(0)=8(\aL-2)$. Hence
\begin{equation}
\lim_{a\to0^+}\Psi(a,\aL) = \frac{0}{16}+\frac{8(\aL-1)}{8(\aL-2)}=\frac{\aL-1}{\aL-2}.
\end{equation}
For $\aL\in(0,1)$, the numerator $\aL-1$ and the denominator $\aL-2$ are both negative, so the
limit is strictly positive. It ranges from $0^+$ (as $\aL\to1^-$) to $\frac{1}{2}$ (as
$\aL\to0^+$).

\paragraph{Part (ii): $a\to+\infty$.}
As $a\to\infty$, $u/t\to1$. Dividing by the leading powers of $t$ and using the leading
coefficients in Appendix~\ref{explicit_poly} gives the following limits.
\begin{itemize}
  \item $N(t)/t^6\to-1-1=-2$ and $D(t)/t^8\to2+2=4$. Hence
  $a\,\mathcal{B}_{\text{DD}}=t^2N(t)/D(t)\to-\frac24=-\frac12$.
  \item $N(t,\aL)/t^{12}\to(2\aL-3)+(2\aL-3)=2(2\aL-3)$ and $D(t,\aL)/t^{16}\to2(\aL-2)$. Hence
  $a^2\,\mathcal{B}_{\text{TSRN}}=t^4N(t,\aL)/D(t,\aL)\to(2\aL-3)/(\aL-2)$, so
  $a\,\mathcal{B}_{\text{TSRN}}\to0$.
\end{itemize}
Therefore $a\,\Psi(a,\aL)\to-\frac12$, that is, $\Psi(a,\aL)=-\frac{1}{2a}+o\bigl(\frac1a\bigr)$ as
$a\to\infty$. The coefficient $-\frac12$ does not depend on $\aL$, and $\Psi(a,\aL)\to0^-$ for every
$\aL\in(0,1)$.
\end{proof}

\subsection{\texorpdfstring{Proof of Theorem~\ref{thm:unique}: Uniqueness of the Critical Threshold}{Proof of the uniqueness theorem}}
\label{app:unique_zero}

\begin{proof}
Fix $\aL\in(0,1)$ and write $t=\sqrt a>0$ and $u=\sqrt{a+4}=\sqrt{t^2+4}>0$, so that $u^2=t^2+4$.
Throughout, a prime denotes differentiation with respect to $a$ (or with respect to the first
argument $s$ of $Q$).

\emph{Common denominator.}
By the proofs of Theorems~\ref{thm:dd_under} and~\ref{thm:tsrn_over},
$\mathcal{B}_{\text{DD}}=N(t)/D(t)$ and $\mathcal{B}_{\text{TSRN}}=N(t,\aL)/D(t,\aL)$, with
$D(t)>0$ and $D(t,\aL)<0$ for all $a>0$. All four polynomials are listed in
Appendix~\ref{explicit_poly}. Hence
\begin{equation}\label{eq:Psi_num_den}
\Psi=\frac{\mathrm{Num}}{\mathrm{Den}},\qquad
\mathrm{Num}\coloneqq N(t)\,D(t,\aL)+N(t,\aL)\,D(t),\qquad
\mathrm{Den}\coloneqq D(t)\,D(t,\aL)<0 .
\end{equation}
So the zeros of $\Psi$ in $(0,\infty)$ are exactly the zeros of $\mathrm{Num}$, and $\Psi$ and
$\mathrm{Num}$ have opposite signs.

Each of $N(t)$, $D(t)$, $N(t,\aL)$ and $D(t,\aL)$ has the form $E+Ou$, where $E$ contains only
even powers of $t$ and $O$ only odd powers (Appendix~\ref{explicit_poly}). For two such
factors,
\[
(E_1+O_1u)(E_2+O_2u)=\bigl(E_1E_2+(t^2+4)\,O_1O_2\bigr)+\bigl(E_1O_2+O_1E_2\bigr)u ,
\]
and the first bracket is even in $t$ while the second is odd. Reducing with $u^2=t^2+4$ therefore
gives
\[
\mathrm{Num}=L+Mu,
\]
where $L$ is even and $M$ is odd in $t$, and both are linear in $\aL$. They are listed in
Appendix~\ref{explicit_poly}. Hence $L=\ell(a)$ and $M=t\,\mu(a)$ for polynomials $\ell,\mu$ in $a$.
Since $t$ and $u$ are smooth in $a>0$, so are $\mathrm{Num}$ and its conjugate
$\widehat{\mathrm{Num}}\coloneqq L-Mu$.

\emph{The norm of the numerator.}
Since $M^2u^2=t^2\mu(a)^2(t^2+4)=a(a+4)\,\mu(a)^2$, the product
\[
R\coloneqq\mathrm{Num}\cdot\widehat{\mathrm{Num}}=L^2-M^2u^2=\ell(a)^2-a(a+4)\,\mu(a)^2
\]
is a polynomial in $a$. Expanding it gives
\begin{equation}\label{eq:R_fact}
R=16(a+4)^3\,Q(a,\aL),\qquad Q(s,\aL)=A(\aL)\,s^2+B(\aL)\,s+C(\aL),
\end{equation}
with
\[
A(\aL)=(\aL-2)(\aL-4),\quad B(\aL)=-4(2\aL^2-8\aL+7),\quad C(\aL)=16(\aL-1)^2 .
\]
Since $a+4>0$, every zero of $\Psi$ in $(0,\infty)$ is a positive root of $Q(\cdot,\aL)$.

\emph{The roots of $Q$.}
For $\aL\in(0,1)$ we have the following signs.
\begin{itemize}
  \item $A>0$, as a product of two negative factors.
  \item $B<0$, since $2\aL^2-8\aL+7>0$ on $(0,1)$: its roots $2\pm\frac{\sqrt2}{2}$ lie outside
  this interval.
  \item $C>0$.
\end{itemize}
The discriminant is $\Delta=B^2-4AC=16(8\aL^2-24\aL+17)$. The inner polynomial has roots
$\frac32\pm\frac{\sqrt2}{4}$, both outside $[0,1]$, and it equals $17>0$ at $\aL=0$. Hence
$\Delta>0$ on $(0,1)$. So $Q(\cdot,\aL)$ has two distinct real roots
\[
a_1(\aL)=\frac{-B-\sqrt\Delta}{2A}<a_2(\aL)=\frac{-B+\sqrt\Delta}{2A},
\]
both simple. By Vieta's formulas both are positive, since their sum $-B/A$ and their product $C/A$
are positive. Moreover $Q'(a_i,\aL)=2Aa_i+B=\mp\sqrt\Delta\neq0$, and $Q$ changes sign at each root.

Next, $Q(1,\aL)=A+B+C=9\aL^2-6\aL-4$. This is convex in $\aL$ and takes the values $-4$ at
$\aL=0$ and $-1$ at $\aL=1$, so $Q(1,\aL)<0$. Also $Q(\frac43,\aL)=\frac{64}{9}(\aL^2-1)<0$ and
$Q(4,\aL)=32>0$. Since $A>0$, $Q(\cdot,\aL)$ is negative exactly on $(a_1,a_2)$. It follows that
\begin{equation}\label{eq:root_location}
0<a_1(\aL)<1<\tfrac43<a_2(\aL)<4 .
\end{equation}
Consequently the zeros of $\Psi(\cdot,\aL)$ form a subset of $\{a_1(\aL),a_2(\aL)\}$.

\emph{Every zero of $\Psi$ is simple.}
Let $\Psi(a_0,\aL)=0$, so that $\mathrm{Num}(a_0)=0$ and $a_0\in\{a_1,a_2\}$.

We first show $\widehat{\mathrm{Num}}(a_0)\neq0$. Suppose instead $\widehat{\mathrm{Num}}(a_0)=0$.
Adding and subtracting the two equations gives $L(a_0)=0$ and $M(a_0)u(a_0)=0$. Since $t,u>0$,
this means $\ell(a_0)=\mu(a_0)=0$. Then
\[
R'(a_0)=2\ell(a_0)\ell'(a_0)-(2a_0+4)\,\mu(a_0)^2-2a_0(a_0+4)\,\mu(a_0)\mu'(a_0)=0 .
\]
But~\eqref{eq:R_fact} gives
\[
R'(a_0)=48(a_0+4)^2\,Q(a_0,\aL)+16(a_0+4)^3\,Q'(a_0,\aL)=16(a_0+4)^3\,Q'(a_0,\aL)\neq0,
\]
because $a_0$ is a simple root of $Q$. This is a contradiction.

Differentiating $R=\mathrm{Num}\cdot\widehat{\mathrm{Num}}$ at $a_0$ and using
$\mathrm{Num}(a_0)=0$ gives
\[
R'(a_0)=\mathrm{Num}'(a_0)\,\widehat{\mathrm{Num}}(a_0),\qquad\text{so}\qquad
\mathrm{Num}'(a_0)=\frac{R'(a_0)}{\widehat{\mathrm{Num}}(a_0)}\neq0 .
\]
Since $\mathrm{Num}(a_0)=0$, the quotient rule gives
$\Psi'(a_0)=\mathrm{Num}'(a_0)/\mathrm{Den}(a_0)\neq0$. So every zero of $\Psi(\cdot,\aL)$ is
simple, and $\Psi$ changes sign at each of its zeros.

\emph{Uniqueness.}
$\Psi(\cdot,\aL)$ is continuous on $(0,\infty)$, has at most two zeros, and changes sign at each of
them. By Lemma~\ref{lem:asymp}, it is positive near $0$ and negative near $\infty$. So it has an
odd number of zeros, hence exactly one. Call it $a^*(\aL)$. Then $\Psi>0$ on $(0,a^*(\aL))$ and
$\Psi<0$ on $(a^*(\aL),\infty)$. Moreover $a^*(\aL)$ equals exactly one of $a_1(\aL)$ and
$a_2(\aL)$, since $a_1\neq a_2$ and $\Psi$ has only one zero.

\emph{Identifying the genuine root at $\aL=\frac12$.}
We evaluate $\Psi(1,\frac12)$ exactly from the polynomials in Appendix~\ref{explicit_poly}. At
$a=1$ we have $t=1$ and $u=\sqrt5$.
\begin{itemize}
  \item \emph{DD term.} $N(1)=-10-4\sqrt5$ and $D(1)=180+80\sqrt5=20(9+4\sqrt5)$. Since
  $(9+4\sqrt5)(9-4\sqrt5)=1$,
  \[
  \mathcal{B}_{\text{DD}}(1)=-\frac{(10+4\sqrt5)(9-4\sqrt5)}{20}=-\frac{5-2\sqrt5}{10}.
  \]
  \item \emph{TSRN term.} The values at $t=1$ are $p_1=890$, $q_1=398$, $p_0=-1165$, $q_0=-521$,
  $r_1=4215$ and $s_1=1885$. So $N(1,\frac12)=-720-322\sqrt5$ and
  $D(1,\frac12)=-\frac32(4215+1885\sqrt5)=-\frac{15}{2}(843+377\sqrt5)$. Since
  $(843+377\sqrt5)(843-377\sqrt5)=843^2-5\cdot377^2=4$,
  \[
  \mathcal{B}_{\text{TSRN}}(1,\tfrac12)=\frac{(720+322\sqrt5)(843-377\sqrt5)}{30}
  =\frac{6\sqrt5-10}{30}=\frac{3\sqrt5-5}{15}.
  \]
\end{itemize}
Therefore
\[
\Psi(1,\tfrac12)=\frac{3\sqrt5-5}{15}-\frac{5-2\sqrt5}{10}=\frac{12\sqrt5-25}{30}>0,
\]
because $(12\sqrt5)^2=720>625=25^2$. By~\eqref{eq:root_location}, $a_1(\frac12)<1$. If
$a^*(\frac12)$ were $a_1(\frac12)$, then $\Psi(\cdot,\frac12)$ would be negative on
$(a_1(\frac12),\infty)$, which contains $1$. This contradicts $\Psi(1,\frac12)>0$. Hence
$a^*(\frac12)=a_2(\frac12)=\frac43+\frac{8\sqrt7}{21}$.

\emph{Extending to all $\aL\in(0,1)$.}
For $i=1,2$ let
\[
S_i\coloneqq\bigl\{\aL\in(0,1):\Psi\bigl(a_i(\aL),\aL\bigr)=0\bigr\} .
\]
The following three facts hold.
\begin{itemize}
  \item The maps $\aL\mapsto a_i(\aL)$ are continuous on $(0,1)$, since $A>0$ and $\Delta>0$
  there.
  \item $\Psi=\mathcal{B}_{\text{DD}}+\mathcal{B}_{\text{TSRN}}$ is jointly continuous on
  $(0,\infty)\times(0,1)$, as a sum of quotients of continuous functions of $(a,\aL)$ whose
  denominators $D(t)$ and $D(t,\aL)$ do not vanish.
  \item By the uniqueness step, $(0,1)=S_1\cup S_2$ with $S_1\cap S_2=\emptyset$.
\end{itemize}
By the first two facts, $\aL\mapsto\Psi(a_i(\aL),\aL)$ is continuous on $(0,1)$, so both $S_i$ are
closed in $(0,1)$. By the third fact, each $S_i$ is also open in $(0,1)$. Since $(0,1)$ is
connected and $\frac12\in S_2$, it follows that $S_2=(0,1)$. Therefore $a^*(\aL)=a_2(\aL)$ for
all $\aL\in(0,1)$.

\emph{Closed-form expression.}
Hence
\begin{equation}\label{eq:astar}
a^*(\aL)=a_2(\aL)=\frac{-B(\aL)+\sqrt{\Delta(\aL)}}{2A(\aL)}
=\frac{4(2\aL^2-8\aL+7)+4\sqrt{8\aL^2-24\aL+17}}{2(\aL-2)(\aL-4)} .
\end{equation}
At $\aL=\frac12$ this gives $a^*=\frac43+\frac{8\sqrt7}{21}\approx2.34$.

\emph{Sign characterization.}
By the uniqueness step, $\Psi>0$ on $(0,a^*(\aL))$ and $\Psi<0$ on $(a^*(\aL),\infty)$. By
Theorems~\ref{thm:dd_under} and~\ref{thm:tsrn_over},
$\mathcal{B}_{\text{DD}}<0<\mathcal{B}_{\text{TSRN}}$, so
$\Psi=|\mathcal{B}_{\text{TSRN}}|-|\mathcal{B}_{\text{DD}}|$. Hence
$|\mathcal{B}_{\text{DD}}|<|\mathcal{B}_{\text{TSRN}}|$ for $a<a^*(\aL)$, and
$|\mathcal{B}_{\text{DD}}|>|\mathcal{B}_{\text{TSRN}}|$ for $a>a^*(\aL)$, extending to the corresponding
statements about $|\BiasDD(\eps)|$ and $|\BiasTSRN(\eps)|$ for small $\eps>0$.
\end{proof}

\subsection{Proof of Theorem~\ref{thm:mono}: Monotonicity of \texorpdfstring{$a^*(\aL)$}{a*(alphaL)}}
\label{app:monotonicity}

\begin{proof}
By the proof of Theorem~\ref{thm:unique}, $a^*(\aL)=a_2(\aL)=\bigl(-B+\sqrt\Delta\bigr)/(2A)$, the
larger root of $Q(\cdot,\aL)$. Since $A>0$ and $\Delta>0$ on $(0,1)$, $a^*$ is continuously
differentiable on $(0,1)$. Differentiating the identity $Q(a^*(\aL),\aL)=0$ gives
\begin{equation}\label{eq:IFT_da}
\frac{\partial Q}{\partial s}\bigl(a^*,\aL\bigr)\,\frac{\dd a^*}{\dd\aL}
+\frac{\partial Q}{\partial\aL}\bigl(a^*,\aL\bigr)=0 .
\end{equation}

$\frac{\partial Q}{\partial s}(a^*,\aL)=2A\,a^*+B=\sqrt{\Delta}>0$.

Differentiating $A$, $B$ and $C$ gives
\[
\frac{\partial Q}{\partial\aL}(s,\aL)=(2\aL-6)\,s^2+(32-16\aL)\,s+32(\aL-1)
=2(s-4)\bigl[(\aL-3)\,s+4(1-\aL)\bigr].
\]
By~\eqref{eq:root_location}, $\frac43<a^*<4$, so $a^*-4<0$. Since $\aL-3<0$ and $a^*>\frac43$,
\[
(\aL-3)\,a^*+4(1-\aL)<(\aL-3)\tfrac43+4(1-\aL)=-\tfrac{8}{3}\aL<0 .
\]
Hence $\frac{\partial Q}{\partial\aL}(a^*,\aL)>0$.

By~\eqref{eq:IFT_da},
$\dd a^*/\dd\aL=-\frac{\partial Q}{\partial\aL}(a^*,\aL)\big/\sqrt{\Delta}<0$ for all
$\aL\in(0,1)$.
\end{proof}

\section{Additional Experimental Results}
\label{app:experiments}

\subsection{Complete Scenario Specifications}

Table~\ref{tab:full_scenarios} provides an insight into all 40 test scenarios.

{
\footnotesize
\begin{longtable}{cl}
\caption{Simulation Scenarios and Descriptions}
\label{tab:full_scenarios} \\
\toprule
\textbf{Scenario} & \textbf{Description} \\
\midrule
\endfirsthead

\toprule
\textbf{Scenario} & \textbf{Description} \\
\midrule
\endhead

\midrule
\multicolumn{2}{r}{\emph{Continued on next page}} \\
\endfoot

\bottomrule
\endlastfoot
S1  & Uniform improvement: $+0.1$ to all $P_{ij}$. \\
S2  & Single cell change: $P_{00}$ increases from $0.1$ to $0.5$. \\

S3  & Treatment helps one listing type and hurts another. \\

S4  & Small heterogeneous effect by listing type . \\

S5  & Small heterogeneous effect by customer type. \\

S6  & High supply, low demand ($\tau=2.0$, $\mu=0.1$). \\

S7  & Low supply, high demand ($\tau=0.1$, $\mu=2.0$). \\

S8  & Many customer types (10 customers, 2 listings). \\

S9  & Many listing types (2 customers, 10 listings). \\

S10 & Sparse market with specialized customer--listing matches. \\

S11 & Dense market with uniform moderate matching probabilities. \\

S12 & One dominant customer type (90\% mass on type 0). \\

S13 & One dominant listing type (90\% mass on listing 0). \\

S14 & Treatment primarily benefits rare listing types. \\

S15 & Treatment primarily benefits rare customer types. \\

S16 & Asymmetric treatment split ($a_C=0.2$, $a_L=0.8$). \\

S17 & Very small uniform treatment effect ($+0.01$). \\

S18 & Very large treatment effect (mean $P$ increases from $\sim0.15$ to $\sim0.75$). \\

S19 & Heterogeneous listing turnover rates ($\tau=[0.1, 2.0]$). \\

S20 & Heterogeneous customer outside options ($\epsilon=[0.05, 0.2, 0.8]$). \\

S21 & Treatment hurts outcomes (negative global treatment effect). \\

S22 & Very low outside option: captive customers ($\epsilon \approx 0.01$). \\

S23 & Interaction effects: treatment varies by customer--listing pair. \\

S24 & Column permutation: listing preferences are swapped. \\

S25 & Extreme scarcity ($\tau=0.05$, $\mu=5.0$). \\

S26 & Treatment affects diagonal matches only (specialization). \\

S27 & Treatment affects off-diagonal matches only (cross-matching). \\

S28 & Large market ($5 \times 5$) with uniform additive effect. \\

S29 & Power-law customer distribution. \\

S30 & Power-law listing distribution. \\

S31 & Multiplicative treatment effect ($P \mapsto 1.3P$). \\

S32 & Only weak matches ($P<0.5$) improve. \\

S33 & Only strong matches ($P \ge 0.5$) improve. \\

S34 & Single listing type ($m=1$). \\

S35 & Single customer type ($n=1$). \\

S36 & Extreme supply constrained ($\tau=0.02$, $\mu=10.0$). \\

S37 & Extreme demand constrained ($\tau=5.0$, $\mu=0.01$). \\

S38 & Outside option dominates utility ($\epsilon$ large). \\

S39 & Treatment creates new matches ($P_{ij}: 0 \rightarrow >0$). \\

S40 & Cannibalization effect from premium listing promotion. \\

\end{longtable}
}

\subsection{One-Sided vs Two-Sided Randomization}

We include the one-sided CR estimator in this figure as well.

\begin{figure}[h!]
\centering
\includegraphics[width=.7\textwidth]{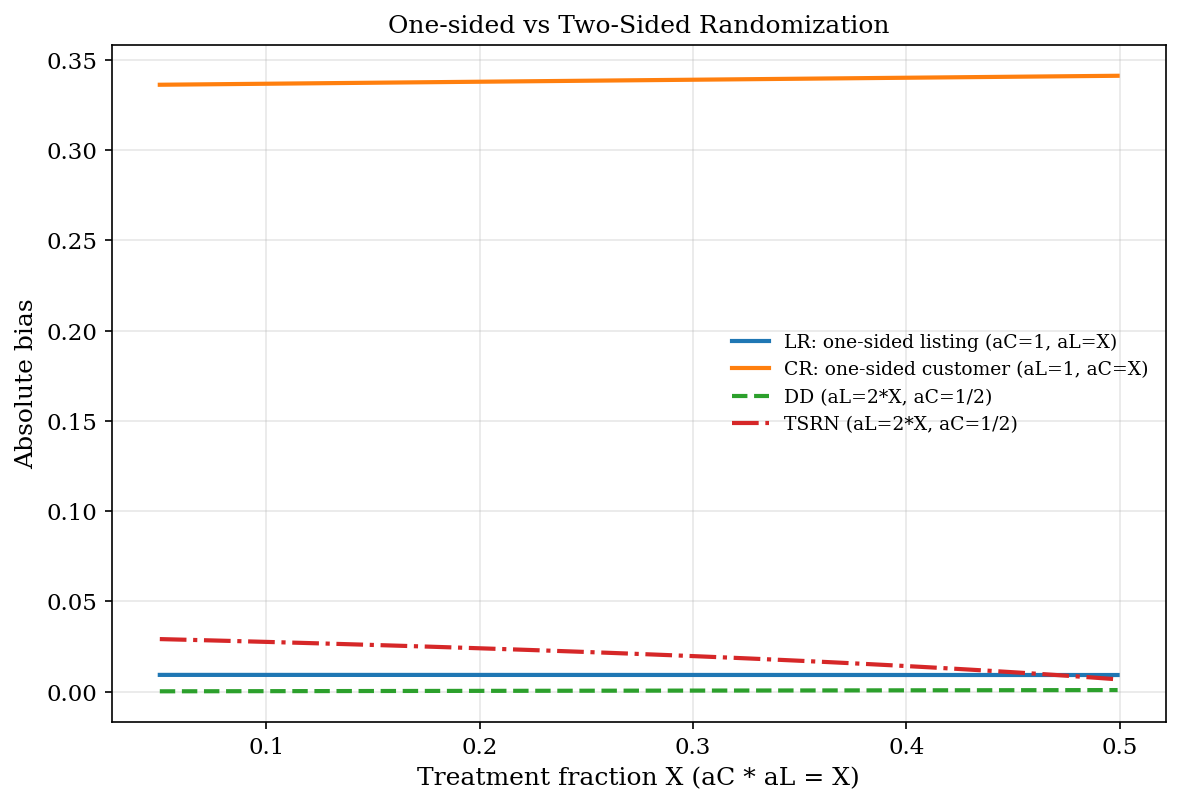}
\caption{Bias across one sided randomizations for LR and CR and two sided for DD and TSR. DD achieves minimum bias consistently.}
\label{fig:onetwoappendix}
\end{figure}

\subsection{Experimental Summary with \texorpdfstring{$\aC=\aL=0.2$}{alphaC = alphaL = 0.2}}

Here, we present our experimental summary for the setting where $\aC=\aL=0.2$, which might be a more realistic setting for online marketplaces.

\begin{table}[H]
\caption{Summary statistics across 40 scenarios.}
\label{tab:summaryappendix}
\centering
\begin{tabular}{lcccccc}
\toprule
& DD & TSRN & CR & LR & TSRI1 & TSRI2\\
\midrule
Lowest $|\text{bias}|$ (count) & \textbf{36/40} & 0/40 & 0/40 & 0/40 & 0/40 & 4/40\\
Median $|\text{rel. bias}|$ (\%) & \textbf{66.5} & 189.3 & 211.5 & 214.8 & 196.1 & 178.0 \\
$90^{th}$ percentile  $|\text{rel bias}|$ (\%) & \textbf{85.0} & 1303.7 &  1531.1 & 1443.0 & 1329.8 & 1226.8\\
Scenarios with $|\text{rel bias}|<$ 25\% & \textbf{2/40} & 1/40 & 0/40 & 0/40 & 1/40 & 1/40\\
Scenarios with $|\text{rel bias}|>$ 100\% & \textbf{3/40} & 36/40 &  36/40 & 36/40 & 36/40 & 36/40\\
\bottomrule
\end{tabular}
\end{table}

\section{Explicit Polynomials}
\label{explicit_poly}

In this section, we present explicit forms of polynomials involved in the proofs, with $t=\sqrt a$, $u=\sqrt{t^2+4}$ and
$s=t^2=a$. These forms can also be found in the code. In every case the $u$-free part is even in
$t$ and the coefficient of $u$ is odd in $t$.

\paragraph{Theorem~\ref{thm:dd_under}.}
$\mathcal{B}_{\text{DD}}=N(t)/D(t)$ with $D(t)=d_0(t)+d_1(t)\,u$, where
\begin{align*}
N(t)&=-t^6-5t^4-4t^2-(t^5+3t^3)\,u,\\
d_0(t)&=2t^8+20t^6+66t^4+76t^2+16,\\
d_1(t)&=2t^7+16t^5+38t^3+24t .
\end{align*}

\paragraph{Theorem~\ref{thm:tsrn_over}.}
$\mathcal{B}_{\text{TSRN}}=N(t,\aL)/D(t,\aL)$ with
$N(t,\aL)=\aL\,[p_1(t)+q_1(t)\,u]+[p_0(t)+q_0(t)\,u]$ and
$D(t,\aL)=(\aL-2)\,[r_1(t)+s_1(t)\,u]$, where
\begin{align*}
r_1(t)&=t^{16}+18t^{14}+133t^{12}+518t^{10}+1134t^8+1372t^6+833t^4+198t^2+8,\\
s_1(t)&=t^{15}+16t^{13}+103t^{11}+340t^9+606t^7+560t^5+231t^3+28t,\\
p_1(t)&=2t^{12}+26t^{10}+128t^8+294t^6+310t^4+122t^2+8,\\
q_1(t)&=2t^{11}+22t^9+88t^7+154t^5+110t^3+22t,\\
p_0(t)&=-3t^{12}-38t^{10}-181t^8-398t^6-395t^4-142t^2-8,\\
q_0(t)&=-3t^{11}-32t^9-123t^7-204t^5-135t^3-24t .
\end{align*}
At $\aL=1$,
\begin{align*}
p_1+p_0&=-t^{12}-12t^{10}-53t^8-104t^6-85t^4-20t^2,\\
q_1+q_0&=-t^{11}-10t^9-35t^7-50t^5-25t^3-2t .
\end{align*}

\end{document}